\documentclass{aa}  

\usepackage{graphicx}
\usepackage{txfonts}
\newcommand{\sub}[1]{_{\rm #1}}
   
\newcommand{\HI}{H\,{\sc i}}

\newcommand{\changed}{}
\newcommand{\changedtwo}{}
\newcommand{\nonprinted}[1]{}

\begin{document}

   \title{The reliability of observational measurements of column density probability distribution functions}

	\titlerunning{Observational measurements of column density PDFs}

   \author{V. Ossenkopf-Okada    \inst{1} \and
          T. Csengeri     \inst{2} \and
          N. Schneider \inst{1} \and
          C. Federrath    \inst{3} \and
          R.S. Klessen    \inst{4}
          }

   \institute{I.\,Physikalisches Institut, Universit\"at zu K\"oln, 
   	Z\"ulpicher Stra{\ss}e 77, 50937 K\"oln, Germany\\
              \email{ossk@ph1.uni-koeln.de}
         	            \and
             Max-Planck Institut f\"ur Radioastronomie, Auf dem H\"ugel 69, Bonn, Germany
             \and
             Research School of Astronomy and Astrophysics, Australian National University, Canberra, ACT 2611, Australia
              \and
              Universit\"at Heidelberg, Zentrum f\"ur Astronomie, Institut f\"ur Theoretische Astrophysik, 69120 Heidelberg, Germany
             }

   \date{Received ; accepted }

% \abstract{}{}{}{}{} 
% 5 {} token are mandatory
 
  \abstract
  % context heading (optional)
  % {} leave it empty if necessary  
   {Probability distribution functions (PDFs) of column densities
   are an established tool to characterize the evolutionary state of interstellar
   clouds.}
  % aims heading (mandatory)
   {Using simulations, we show to what degree their determination 
   is affected by  noise, line-of-sight contamination, field selection, 
   and the incomplete sampling in interferometric measurements.}
  % methods heading (mandatory)
   {We solve the integrals that describe the convolution of a 
   cloud PDF with contaminating sources such as noise and 
   line-of-sight emission and study the impact of missing
   information on the measured column density PDF. In this way we can quantify 
   the effect of the different processes and propose ways to correct 
   for their impact in order to recover the intrinsic PDF of the observed
   cloud.}
  % results heading (mandatory)
   {The effect of observational noise can be easily estimated and corrected
   for if the root mean square (rms) of the noise is known. 
   For {\changed $\sigma\sub{noise}$ values below 40\,\% of the typical cloud column 
   density, $N\sub{peak}$,} this involves almost no degradation of the
   accuracy of the PDF parameters. For higher noise levels and narrow
   cloud PDFs the width of the PDF becomes increasingly uncertain. 
   A contamination by turbulent foreground or background clouds can be 
   removed as a constant shield if the PDF of the contamination peaks at a
   lower column or is narrower than that of the observed cloud. 
   Uncertainties in the definition of the cloud boundaries mainly 
   affect the low-column density part of the PDF and
   the mean density. As long as more than 50\,\% of a cloud are 
   covered, the impact on the PDF parameters is negligible.
   In contrast, the incomplete sampling of the uv plane in 
   interferometric observations leads to uncorrectable distortions of 
   the PDF of the produced maps. An extension of ALMA's capabilities
   would allow us to recover the high-column density tail of the
   PDF but we found no way to measure the intermediate and
   low column density part of the underlying cloud PDF in 
   interferometric observations.
   }
  % conclusions heading (optional), leave it empty if necessary 
   {}


   \maketitle
%
%________________________________________________________________

\section{Introduction}

Probability distribution functions (PDFs) of column density maps
of interstellar clouds have been proven to be a valuable tool to
characterize the properties and evolutionary state of the clouds
(see e.g. \cite{BerkhuijsenFletcher2008,
Kainulainen2009, FroebrichRowles2010, Lombardi2011, Kainulainen2011,
Schneider2012, Schneider2013, Russeil2013, KainulainenTan2013, 
Hughes2013, Alves2014, Druard2014, Schneider2015a, Schneider2015b,
Schneider2015d, BerkhuijsenFletcher2015} for the observational side
and \cite{Klessen2000,
Federrath2008, Federrath2010, Kritsuk2011, Molina2012, Konstandin2012,
FederrathKlessen2013, Girichidis2014} for the theoretical side).
Column density maps of dense clouds typically
show a log-normal PDF at low column densities, produced by interstellar
turbulence \citep{VazquezSemadeni1994,Padoan1997,PassotVazquezSemadeni1998,
Federrath2008, Price2011, FederrathBanerjee2015, Nolan2015}, 
and a power-law tail in the PDF at 
high densities due to gravitational collapse \citep{Klessen2000,
Kritsuk2011, FederrathKlessen2013, Girichidis2014}.
The PDFs seen in 2-D projection allow to estimate the underlying 
3-D {\changed volume} density PDFs when assuming global isotropy 
\citep{Brunt2010a,Brunt2010b,Ginsburg2013, Kainulainen2014}.
The width of the log-normal part allows to quantify the turbulent
driving of the interstellar medium \citep{NordlundPadoan1999,Federrath2008, Federrath2010}. If the driving is known, PDF width and Mach number can 
be used to constrain the magnetic pressure \citep{PadoanNordlund2011,
Molina2012}. The slope of the PDF power-law tail can be translated into radial
profiles of collapsing clouds when assuming a spherical or cylindrical density
distribution \citep{Kritsuk2011, FederrathKlessen2013}.

PDFs can be determined from any mapped quantity, not
only column densities.
For instance, \citet{Schneider2015b} combined PDFs of dust temperatures
with the column-density PDFs, \citet{Burkhart2013} compared PDFs
of molecular lines with different optical depths, and
\citet{MieschScalo1995,Miesch1999,HilyBlant2008,Federrath2010,Tofflemire2011}
studied PDFs of centroid velocities and velocity increments. 
Our analysis does not make any
assumption on the quantity that is actually analyzed. As a pure
statistical analysis, it can be applied to any kind of map.
{\changed However, directly measured quantities have the advantage 
that the impact of observational noise is easier to quantify.
This favors e.g. intensities over velocity centroids or temperatures.} 
For the sake of convenience in naming the quantities and
in the direct comparison to the frequent computation of PDFs from
column density maps, we {\changed focus on} to the column density as
measured quantity, characterized either in terms of gas columns,
visual extinction, or submm-emission from dust with constant 
temperature. Therefore, we use column density $N$ and $A\sub{V}$ synonymously.

\citet{Lombardi2015} discussed the impact of statistical noise, 
contamination with fore- and background material, and boundary
biases on the PDFs of their near-infrared extinction maps and
concluded that it is impossible to reliably determine the lower
column-density tail of the observed molecular cloud PDFs, leaving 
only the high-column density power-law tail as significant
structure.
For column density maps obtained from Herschel PACS and SPIRE
continuum observations, \citet{Schneider2015a} showed in contrast
that 1) the foreground contamination can be corrected for by assuming
a constant screen of material, 2) the selection of the map 
boundaries only affects column densities well below the column
density peak, and 3) the observational noise in the Herschel maps was
an order of magnitude lower allowing to fully resolve the
log-normal part of the turbulent column-density PDF. {\changedtwo
A recent comprehensive study by \citet{Brunt2015} showed that the
neglection of noise and line-of-sight contamination can lead
to an erroneous interpretation of power-law tails.}

In order to provide a general framework to assess the reliability of 
column density PDFs obtained in observations, we
perform here a parameter study where we simulate different
observational biases and limitations with a variable
strength to show under which conditions interstellar cloud PDFs can
be reliably measured, when they can be recovered using parametric
corrections, and when the observational limitations dominate the result.

In Sect.~2, we introduce the formalism and the simulated data used here.
In Sect.~3, we study the impact of observational noise, Sect.~4
deals with line-of-sight contamination, Sect.~5 with the selection
of map boundaries, and Sect.~6 with the limitations in interferometric
observations. Sect.~7 summarizes our findings.

\section{Mathematical description}

\subsection{PDFs}
\label{sect_pdf_math}

PDFs of column densities $p(N)$ are defined as the probability of 
the column density in a map to fall in the interval $N, N+dN$.
As probabilities they are normalized to unity, i.e.
$\int_0^{\infty} p(N) dN=1$. If we discuss noisy 
quantities or other physical quantities that can become negative, 
the integral should start at $-\infty$. 
Due to the huge range of densities and resulting column densities 
in the interstellar medium, it
is more appropriate to use logarithmic bins, i.e. we switch to 
a logarithmic scale $\eta=\ln(N/N\sub{peak})$, where $N\sub{peak}$
is the most probable column density in logarithmic bins, i.e. the 
peak of the distribution on a logarithmic column density scale. 
The translation between the PDFs on the linear and the logarithmic scale
can be easily computed by
\begin{equation}
p_{\eta}(\eta) d\eta = p_{N}(N) \frac{\partial N}{\partial \eta} d\eta = N p_N(N) d \eta\;.
\label{eq_peta_definition}
\end{equation}

Most molecular cloud observations are characterized by a log-normal
part for the high probability of low densities
\begin{equation}    
p_\eta(\eta)=\frac{1}{\sqrt{2\pi}\sigma_{\eta,{\rm cloud}}}\;
\exp \left( -\frac{\eta^2}{2\sigma_{\eta,{\rm cloud}}^2} \right)\,,
\label{eq_lognormal}
\end{equation}
having typical widths on the logarithmic scale $\sigma\sub{cloud}$ between 0.2 
and 0.5 \citep[see e.g.][]{Kainulainen2009,
Hughes2013,Schneider2015a}, and a power-law tail at higher densities
\begin{equation}    
p(\eta)d\eta=\eta^{-s} d\eta \;,
\end{equation}
where the exponent $s$ varies in observations between about 
$s=1\dots 4$ \citep[see e.g.][]{Schneider2015b, StutzKainulainen2015}.
For a free-fall gravitational collapse creating the tail,
we expect values $s=2\dots 3$, depending on
the geometry of the collapse \citep{Kritsuk2011, FederrathKlessen2013,
Girichidis2014}.

All effects of noise, contamination, and edge selection mainly
change the low-density regime, so that it is sufficient to
concentrate on the log-normal part here. The measurement of the
power-law tail can be affected by the impact of insufficient
spatial resolution and binning. That was already studied in detail by 
\citet{Lombardi2010} and \citet{Schneider2015a}. For our semi-analytic 
studies, focusing on low-density effects, we will therefore
represent molecular clouds by a log-normal PDF, fully described
by the two parameters $N\sub{peak}$ and $\sigma\sub{cloud}$,
here. A power-law tail is only added for the full cloud simulations 
(see Sect.~\ref{sect_testcloud}).

The normalization of the density to the peak of the distribution
on a logarithmic scale $N\sub{peak}(\eta)$, deviates from
the normalization by $\langle N \rangle$ that is often employed in
the literature and that we also used in \citet{Schneider2015a}. In principle,
the choice of the normalization constant does not affect any
of our outcomes. Taking the logarithmic peak is simply
more convenient for log-normal distributions centering them at 
$\eta=0$. For a log-normal distribution we find a fixed
relation between the possible normalization constants. The
average column density is $\langle N\rangle = N\sub{peak}(\eta)
\exp(\sigma_{\eta,{\rm cloud}}^2/2)$ and the peak of the probability
distribution in linear units falls at $N\sub{peak}(N)=N\sub{peak}(\eta)
\exp(-\sigma_{\eta,{\rm cloud}}^2)$. However, we will show in Sect.
\ref{sect_edge_effects} that a normalization by the PDF peak is 
much more stable against selection effects than the mean 
column density. Therefore, we generally recommend the choice of
$N\sub{peak}$ for the column density normalization.

\subsection{Contamination}
\label{sect_contamination}

All effects of contamination by observational noise or
other line-of-sight material will be linear in column density $N$,
but not in $\eta$. At every point in the map we measure a column density
that is given by $N\sub{tot}=N\sub{cloud}+N\sub{contam}$, i.e.
on a logarithmic scale we find $\eta\sub{tot}=\ln({N\sub{cloud}+N\sub{contam}})
-\ln(N\sub{peak})$.
The resulting PDF of a contaminated map then results from the
convolution integral of both distributions on the linear scale
\begin{eqnarray}
p_{\eta}(\eta\sub{tot}) \! & = N\sub{tot} \int_{-\infty}^{\infty} 
\! & p_{N,{\rm contam}}(N) \label{eq_convolution} \\
 && \times \frac{p_{\eta,{\rm cloud}}[\ln(N\sub{tot}-N)-\ln(N\sub{peak})]}{N} dN \;.
\nonumber
\end{eqnarray}
We can assume a normal distribution for the
contamination with {\changed direct measurement noise}\footnote{An 
equivalent approach, defined on the linear scale, was recently
introduced by \citet{Brunt2015}.}
\begin{equation}
p_{N,{\rm contam}}(N)=\frac{1}{\sqrt{2\pi}\sigma\sub{noise}}\;
\exp \left( -\frac{N^2}{2\sigma\sub{noise}^2} \right)
\label{eq_gauss}
\end{equation}
{\changed
For derived quantities or measurements in particular observing modes
the errors can be correlated or non-Gaussian requiring the convolution
with a different noise distribution $p_{N,{\rm contam}}(N)$.
For the contamination by a turbulent foreground
or background cloud we can use a second} log-normal distribution 
\begin{equation}
p_{N,{\rm contam}}(N)=\frac{1}{\sqrt{2\pi}\sigma\sub{contam} \times N}\; 
\exp \left( -\frac{(\ln{N}-\ln{N\sub{contam}})^2}{2\sigma\sub{contam}^2} \right)
\label{eq_contamin}
\end{equation}
where $N\sub{contam}$ denotes the most probable column density of the
contamination on a logarithmic scale. 
We assume that the contaminating structures are 
turbulence-dominated so that they do not show any
self-gravitating cores yet and can be described by a pure
log-normal distribution without power-law tail. Collapsing
parts that produce a power-law tail are easily identified
in the maps as small structures so that they can be separated from
large-scale contaminations.

In principle, it is possible to compute the original cloud
distribution $p_{N, {\rm cloud}}(N)$ from  the measured distribution
$p_{N}(N\sub{tot})$ by inverting the convolution in Fourier
space if the contamination PDF, $p_{N, {\rm contam}}(N)$, is known
\begin{equation}
p_{N, {\rm cloud}}(N)=\mathfrak{F}^{-1}\left(\frac{\mathfrak{F}[p_{N}(N)]}
{\mathfrak{F}[p_{N, {\rm contam}}(N)]} \right)
\label{eq_deconvolution}
\end{equation}
where $\mathfrak{F}$ and $\mathfrak{F}^{-1}$ denote the normal
and inverse Fourier transform. 
However, this operation is inherently unstable. It amplifies noise,
gridding effects and numerical uncertainties, so that it only works
with very well behaving denominators. As Gaussian noise is 
mathematically well characterized we will test this method in Sect.~3.

\subsection{Test data}
\label{sect_testcloud}

   \begin{figure}
   \includegraphics[angle=90,width=\columnwidth]{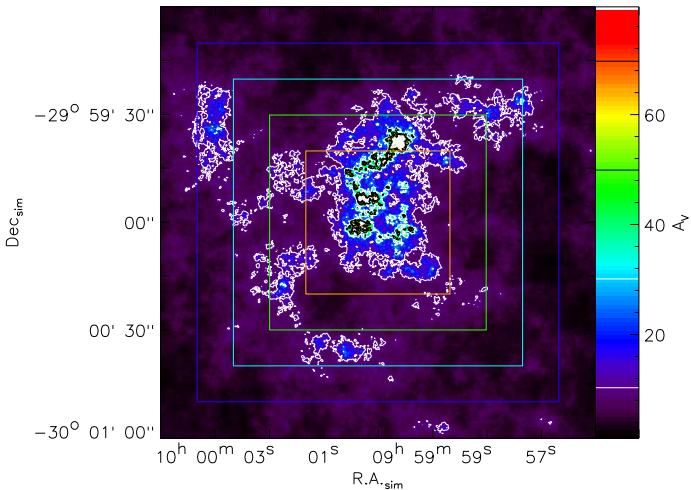}
      \caption{Test data set representing a fractal cloud with an
	ideal column density PDF consisting of a log-normal part
	and a power-law high-density tail. Coordinates are invented to allow
	for an easy comparison with real observations and to perform
	simulated ALMA observations in Sect.~\ref{sect_interferometer}. 
    Contours and colors	visualize the same data. The contour levels 
    are indicated in the color bar. Column densities above
	$A\sub{V}=80$ are saturated in the plot to make the turbulent
    cloud part better visible. The actual maximum column density
	is three times larger.
    As the peak of the log-normal PDF is located
	at $A\sub{V}=5$, the contours almost cover material in the 
    PDF power-law tail only. According to the low probabilities
    in the tail, the fraction of green pixels is already small
    and that of pixels with yellow and red colors is tiny.
    The four colored squares indicate the subregions discussed
    in Sect.~\ref{sect_edge_effects}.}
         \label{fig_fractalmap}
   \end{figure}

For a realistic data set that has all properties of
a molecular cloud column density map, but no observational limitations
we modify a fractional Brownian motion \citep[fBm, see e.g.][]{Stutzki1998}
image to approximate the PDF seen in molecular clouds. 
{\changed The original fBm image is constructed by inverse transform
of a Fourier spectrum that has a given power spectral index $\beta$
and random phases. For the spectral index we use a value} of $\beta=2.8$, 
measured in many molecular clouds \citep[see e.g.][]{Bensch2001,Falgarone2004}
and consistent with numerical simulations \citep{Kowal2007,Federrath2009}.
The fBm {\changed structure reflects the typical spatial correlations 
in interstellar cloud maps. The resulting image always has} a Gaussian PDF 
($p_{N, {\rm fBm}}(N)$ given by Eq. \ref{eq_gauss}). 
To obtain the desired log-normal plus power-law column 
density PDF with a minimum distortion of the spatial
structure we translate the fBm values into cloud column densities by
\begin{equation}
N\sub{cloud} = \Pi\sub{cloud}^{-1}\left[ \Pi\sub{fBm}(N) \right]
\end{equation}
where $\Pi$ denotes the integral over the normalized PDF
\begin{equation}
\Pi(N)=\int_{-\infty}^{N} p_N(N') dN'\;,
\end{equation}
varying between zero and unity. The inversion of the function 
$\Pi^{-1}$ can only be computed numerically.

For the parameters of the PDF we use values measured by
\citet{Schneider2015a} and \citet{Schneider2015d} in 
several star-forming clouds, e.g. Auriga, Orion B, and Cygnus X. 
The peak of the PDF $N\sub{peak}$ is located at $A\sub{V}=5$ and
the width of the log-normal part of the PDF in units of the
natural logarithm of the column density is $\sigma_{\eta,{\rm cloud}}=0.45$.
The transition to the power-law tail occurs at $A\sub{V}=11$ and
the tail has an exponent of $s=1.9$. The PDFs in regions with 
somewhat lower star-forming activity are {\changed shifted} towards smaller 
column densities. \citet{Kainulainen2009,Schneider2015a}
find PDF peaks at $A\sub{V} \approx 2$ and deviation points
to the power law tail at $A\sub{V} = 4\dots 5$. {\changed On} the $\eta$ 
scale, used in all simulations here, this is identical to the 
PDF of our test data, $p_\eta(\eta)$. Only when translating the 
results back to an absolute column density scale, the different 
PDF peak density $N\sub{peak}$ needs to be applied. 
We selected a random seed for the fBm that results in a cloud 
well centered in the map, similar to what an observer might choose.

The resulting map is shown in Fig.~\ref{fig_fractalmap}, where
we arbitrarily placed the map on the Southern sky and assigned
spatial dimensions in such a way, that it would make
sense to observe the cloud with ALMA as simulated in Sect.~6.
The PDF of the cloud is shown in Fig.~\ref{fig_noise_on_full_PDF}.

\section{Noise}
\label{sect_noise}

\subsection{Main impact}

   \begin{figure}
   \includegraphics[angle=90,width=\columnwidth]{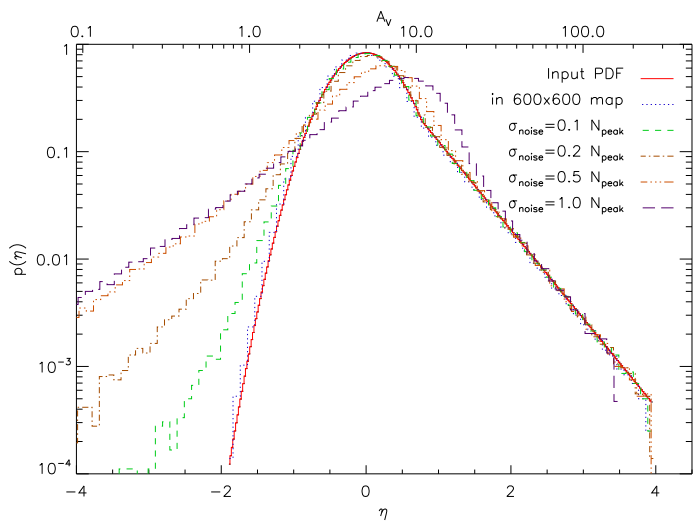}
      \caption{Impact of observational noise on the PDF measured
	for the cloud from Fig.~\ref{fig_fractalmap}. The red solid
	curve shows the analytic description of the input PDF, the
	blue dotted line is the actually measured PDF of the fractal 
    map with a finite pixel number. The other broken lines show 
    the column density
	PDF that is measured if the map is affected by different 
	levels of observational noise, characterized by a normal
	distribution with a standard deviation from 10\,\% to 100\,\% 
    of the peak column density.}
         \label{fig_noise_on_full_PDF}
   \end{figure}

In a first step we study how the PDF is modified by observational
noise, simulated as the superposition of normally distributed random
numbers to the map. The result is shown in Fig.~\ref{fig_noise_on_full_PDF}
where we varied the noise amplitude, i.e. the root-mean-square (rms)
of the noise distribution between 10\,\% and 100\,\% of the peak
density, $N\sub{peak}$, of the cloud PDF.
We see three effects: \\
{\bf i)} An increasing noise level produces a low-density
excess in the PDF, i.e. the log-normal part of the PDF is widened
towards lower densities. If the low densities are completely dominated
by noise, like in the case of noise amplitudes of 50\,\% or more
of the peak column density, we find a linear behavior of
the PDF at low densities $p(\eta) \propto \eta$ due to the factor $N$
in Eq.~\ref{eq_peta_definition} applied to the Gaussian noise
contribution (Eq.~\ref{eq_gauss}) when computing the convolution
integral (Eq.~\ref{eq_convolution}).\\ 
{\bf ii)} For
high noise levels, the additional ``signal'' from the noise also shifts 
the peak of the distribution towards higher column densities, i.e. may
affect the measurement of $N\sub{peak}$.\\ 
{\bf iii)} For the power-law tail towards high column densities, 
however, we find no significant impact. This confirms our earlier
findings \citep{Schneider2015a} and the ones of \citet{Lombardi2015} 
that the log-normal part of the PDF is easily affected by noise, 
but that the power-law tail is stable.

\subsection{Parametric corrections}
\label{sect_noise_correction}

   \begin{figure}
   \includegraphics[angle=90,width=\columnwidth]{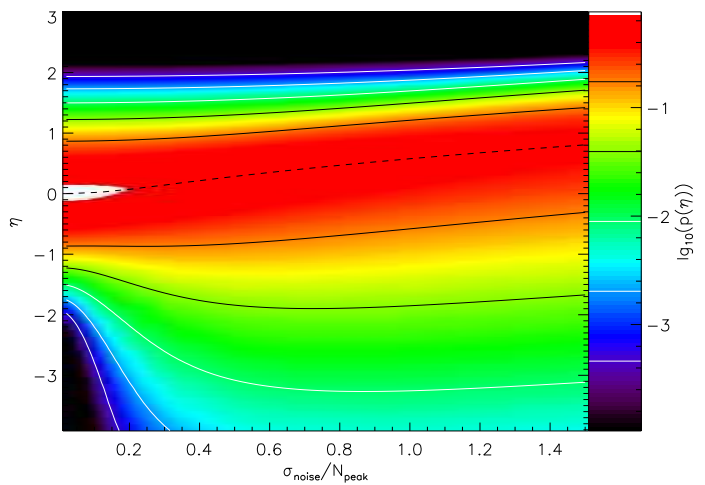}
      \caption{Distributions obtained from the convolution of a log-normal
	cloud PDF (Eq.~\ref{eq_lognormal} with $\sigma_{\eta,{\rm cloud}}=0.5$) with a Gaussian noise PDF
	(Eq.~\ref{eq_gauss}) as a function of the noise amplitude
	($\sigma\sub{noise}$) relative to the peak of the
	log-normal distribution ($N\sub{peak}$). Contours and colors
	visualize the same data, given on a logarithmic scale. The 
    individual contour levels are marked in the color bar. The 
    left edge of the figure represents the original log-normal PDF. 
    When increasing the noise level
	$\sigma\sub{noise}$ in the right direction, we see the distortion
	of the PDF mainly at low column densities $\eta$. To better follow 
	the shift of the PDF maximum to larger column densities with 
    increasing noise level, the dashed line indicates the peak of the
    distribution.}
         \label{fig_pdfscan_noise}
   \end{figure}

A correction of the noise effect for the log-normal part of the PDF is,
however, possible if the amplitude of the noise contamination is known.
To obtain such a correction, we perform a parameter study based on the
analytic description of the PDFs in Eqs.~(\ref{eq_lognormal}-\ref{eq_gauss}),
ignoring the power-law tail that is anyway not sensitive to the noise
contamination. Figure~\ref{fig_pdfscan_noise} shows a first step in such
a parameter scan for the noise amplitude. 
Here, we display all PDFs for the different noise levels
in a color-contour plot. To represent the PDFs in the two-dimensional
plot equivalent to the curves in Fig.~\ref{fig_noise_on_full_PDF}
we show the decadic logarithm of $p_\eta(\eta)$ in colors and
contour levels. The PDFs are computed from the 
convolution integral (Eq.~\ref{eq_convolution}) using a log-normal
cloud PDF with $\sigma_{\eta,{\rm cloud}}=0.5$ and Gaussian noise 
(Eq.~\ref{eq_gauss}) with a varying amplitude.  The left edge of the
figure represents the original log-normal PDF. With increasing noise level
$\sigma\sub{noise}$ towards the right part of the plot, we see the increasing
distortion of each part of the PDF. This allows us to quantify
and correct the distortions based on the noise level. Even for 
low noise levels there
is a low-column-density excess leading to a widening of the PDF. It
stays relatively constant for noise levels $\sigma_{\eta,{\rm cloud}} \ga 
0.5 N\sub{peak}$. For noise amplitudes of $\ga 0.2 N\sub{peak}$ the 
PDF peak is also shifted towards higher column densities.

As the noisy PDFs are no longer log-normal they are no longer 
characterized by two parameters only. We have to discriminate 
between different methods that can 
be used for measuring width and peak position in PDFs. The
problem is similar to the measurement of the position and width of
individual lines measured in a spectroscopic observation. The
use of the absolute peak for the position is sensitive to the 
details of the binning. It can be affected by fluctuations
due to the low-number sampling in individual bins for a
finite map size. This becomes significant for the submaps that
we consider in Sect.~\ref{sect_edge_effects}. 
In contrast, the moments of the distribution, being independent
of the selected binning, are strongly affected by the structure 
of the wings. This would
prevent us from applying the numbers determined here, for the 
contamination of a purely log-normal distribution, to a molecular 
cloud structure that shows an additional power-law tail. 
The compromise, that is often used in the case of spectroscopic 
observations, is a Gaussian fit to the distribution, i.e. a
log-normal fit on our logarithmic $\eta$ scale. For a 
log-normal distribution, all three approaches provide the same 
numbers, but for the noise-contaminated distributions they can 
deviate from each other.

   \begin{figure}
   \includegraphics[angle=90,width=\columnwidth]{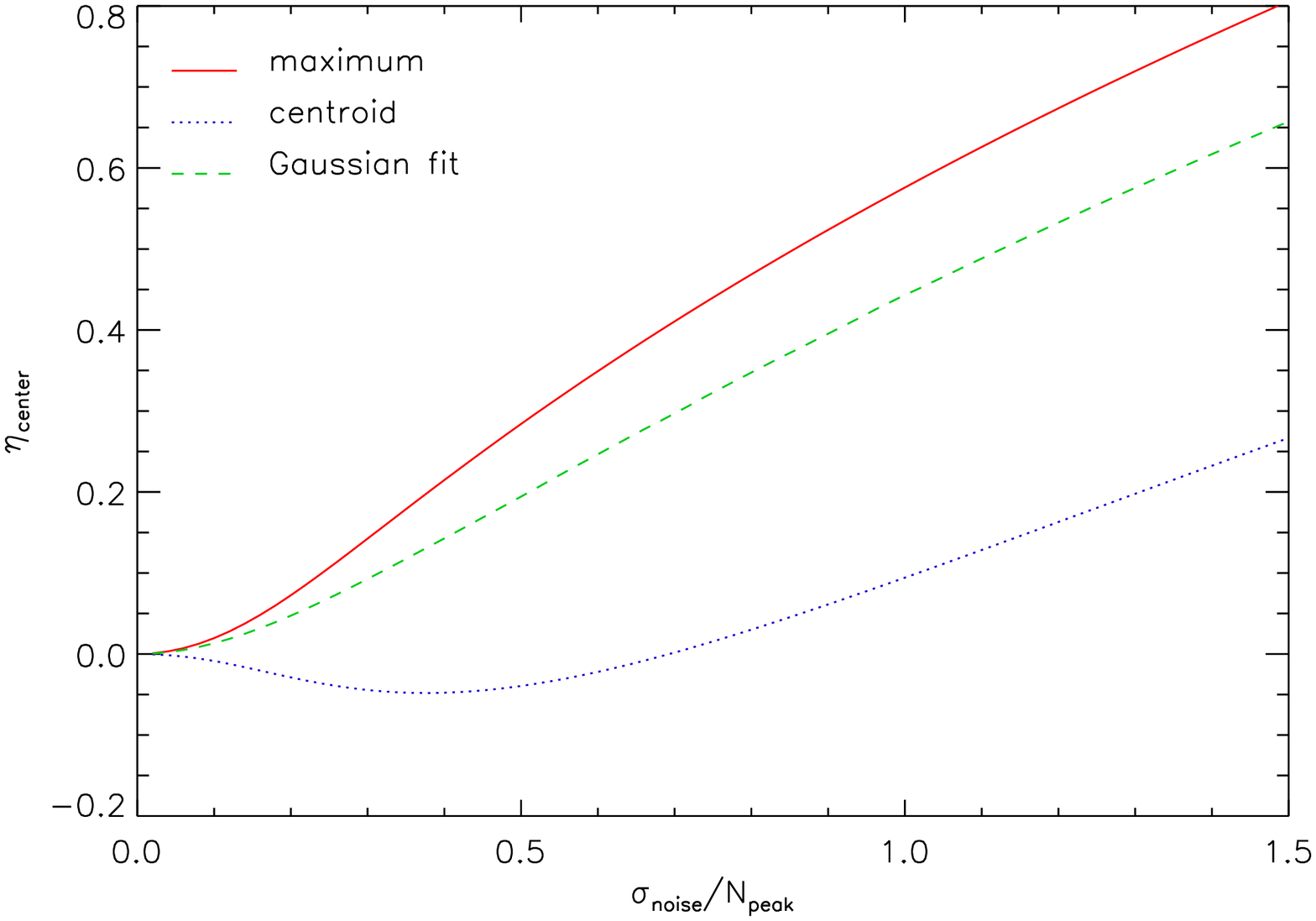}
   \vspace{0.3cm}\\
   \includegraphics[angle=90,width=\columnwidth]{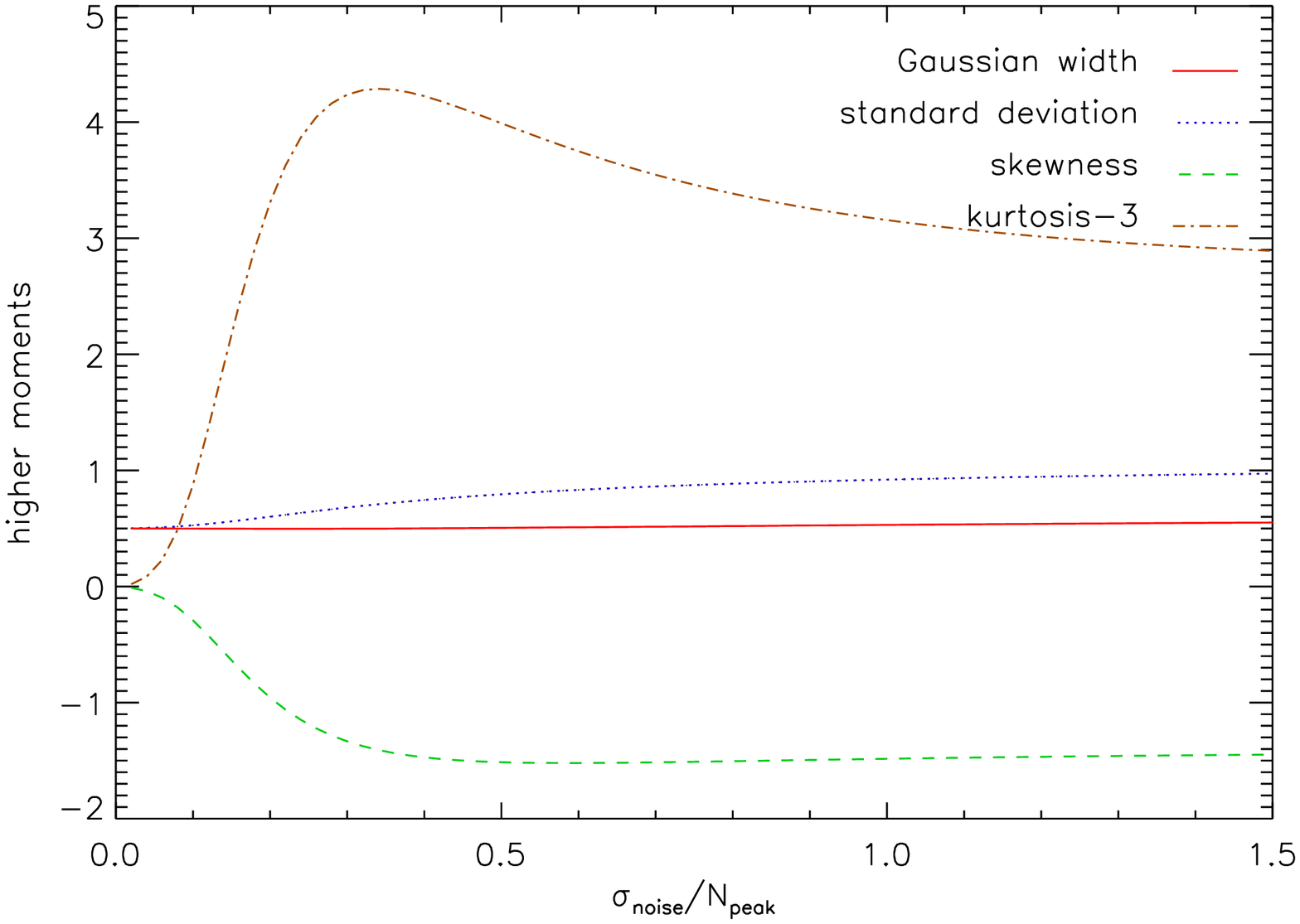}
      \caption{Change of the measured PDF parameters as a function of the 
	noise amplitude. The upper plot shows the measured center position
    using three different measures for the position.
	The red solid line shows the position of the maximum probability,
	the blue dotted line shows the first moment of the distribution, 
	and the green dashed line gives the center of a Gaussian fit
	(in $\eta$). The lower plot compares the standard deviation with
    the width of the Gaussian fit, also showing the higher moments.
    }
         \label{fig_peakscan_noise}
   \end{figure}

Figure~\ref{fig_peakscan_noise} shows the PDF parameters
determined in three different ways as a function of the
amplitude of the noise contamination. The upper plot gives the center
position measured through the maximum, the first moment, and
as the center of a log-normal fit. They agree at zero noise 
contamination. The position of the peak probability gives the highest 
column density values for the PDF center, matching the dashed line in 
Fig.~\ref{fig_pdfscan_noise}.
When considering  the dotted curve for the PDF centroid, we find 
a decrease for the center of the distribution at low noise levels 
compared to the noise-free case due to the widening of the distribution
towards small column densities. The center of a log-normal fit shows an
intermediate behavior. 

The lower plot shows the measured widths of the distribution
together with the next higher moments\footnote{
The third moment of a distribution is the skewness defined
as $S=1/N \times \sum (\eta-\langle \eta \rangle)^3/\sigma^3$,
the fourth moment the kurtosis 
$K=1/N \times \sum (\eta-\langle \eta \rangle)^4/\sigma^4$ where
the sums run over all $N$ pixels of a map and $\sigma$
denotes the standard deviation of the distribution in $\eta$.}. 
The width of a log-normal
fit to the distribution is very stable. It increases by 10\,\%
only. The standard deviation $\sigma$ grows much more, by almost a factor 
of 2. With the formation of the low-density tail due to the
noise contamination, the skewness of the distribution quickly 
drops to about $-1.5$ and the kurtosis grows. Once the
linear dependence seen in Fig.~\ref{fig_noise_on_full_PDF} 
is reached, the skewness 
remains almost constant and only the steepening of the high-density
tail with the shift of the PDF peak is traced through a reduction
of the kurtosis. As the Gaussian fit is least affected by the
formation of the noise tail and it is also applicable for distributions
with a power-law high-density tail, we will continue with this 
log-normal fit characterization for more extended parameter scans.

   \begin{figure}
   \includegraphics[angle=90,width=\columnwidth]{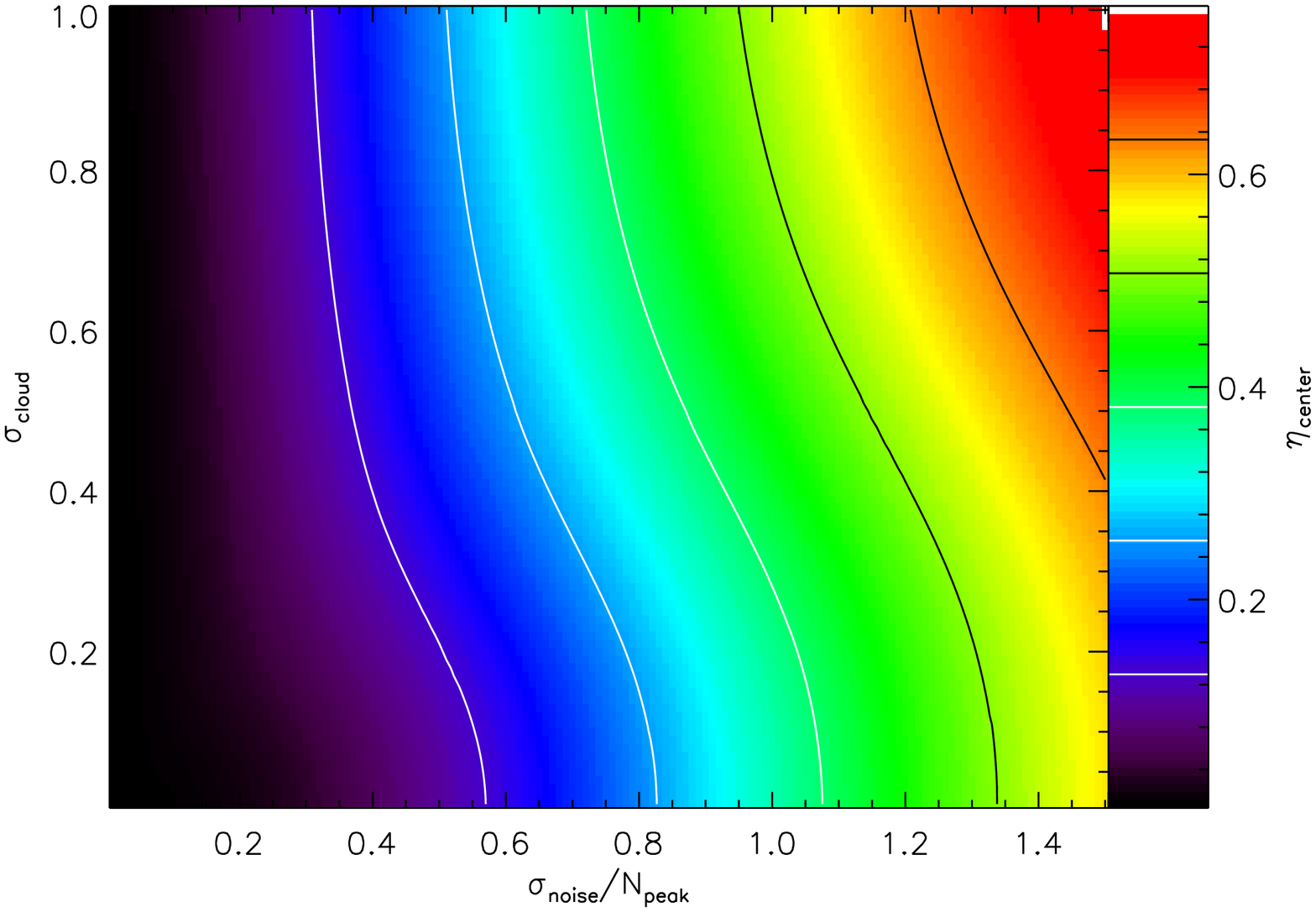}
	\vspace{0.3cm}\\
   \includegraphics[angle=90,width=\columnwidth]{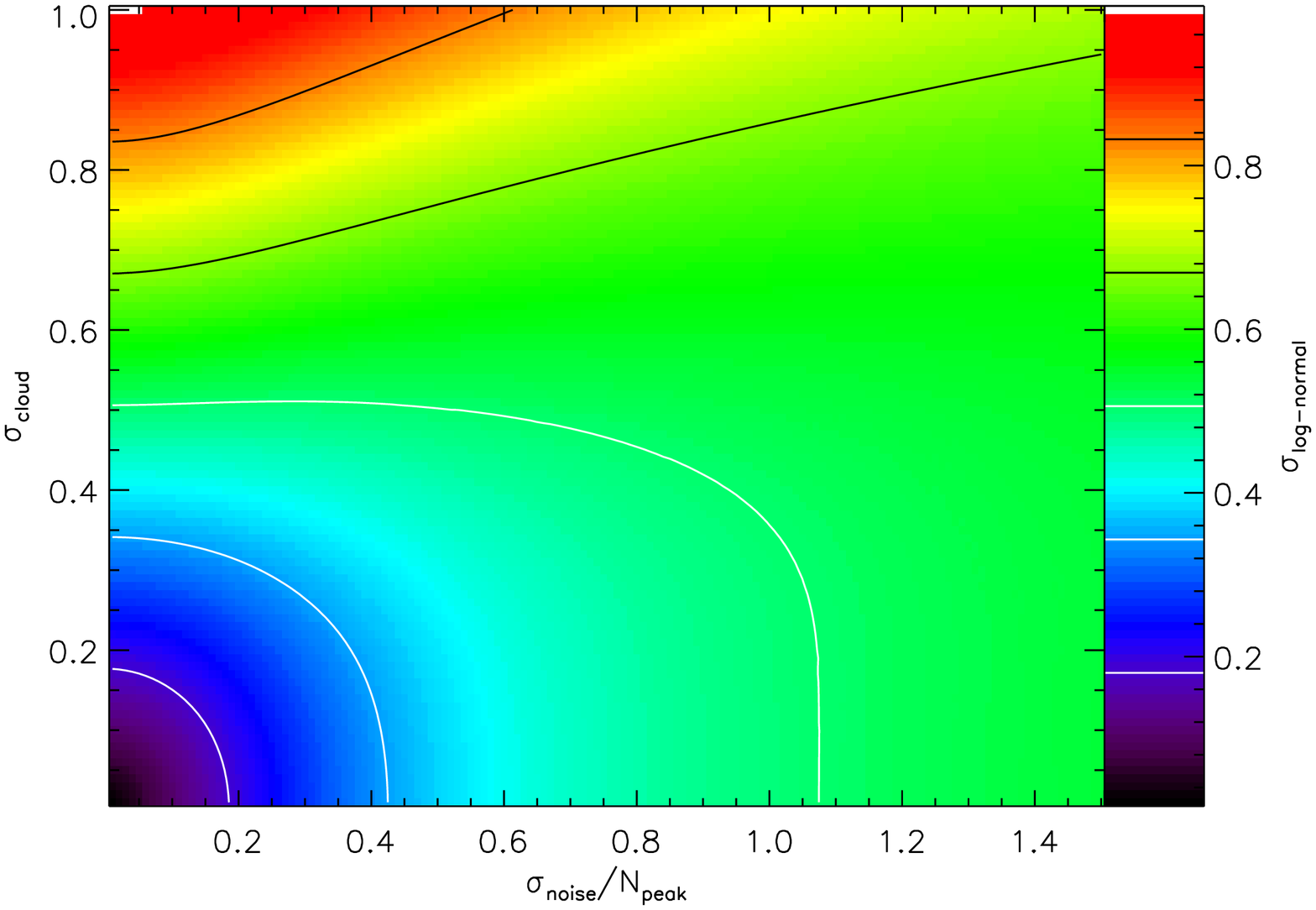}
      \caption{Variation of the parameters of a log-normal fit 
      to the PDFs of noise-contaminated maps as a function of
	noise amplitude and the width of the original cloud PDF.
	The upper plot shows the {\changed measured location of the 
	PDF center}, the lower
	plot the width of the log-normal fit. Without noise contamination
    $\eta\sub{center}=\eta\sub{peak}=0$ and the measured width matches
    the PDF width of the cloud, $\sigma\sub{log-normal}=\sigma\sub{cloud}$. 
    }
         \label{fig_fullnoisescan}
   \end{figure}

When assuming a log-normal cloud PDF, the noise contamination problem is
fully described by two parameters, the width of the cloud PDF and the 
amplitude of the noise distribution relative to the peak of the cloud
PDF. Linear scaling to any cloud observation is straightforward. 
This allows us to provide a complete picture for the correction of noise 
effects. Figure~\ref{fig_fullnoisescan} shows the {\changed new 
PDF peak position and the change of the PDF width}, determined through 
the log-normal fit, as a function of the noise contamination
amplitude and the width of the original cloud PDF. {\changed By definition
the input cloud always has $\eta\sub{center}=\eta\sub{peak}=0$.} For all cloud
distributions we find an almost linear increase of the observed peak 
column density with noise amplitude. The highest shift of the
peak position occurs for broad cloud distributions. The fitted PDF
width reflects the properties of the underlying cloud only for
noise amplitudes below about 0.4-0.5 $N\sub{peak}$. 
{\changed We find two opposite cases for the measured PDF width.
For $\sigma\sub{cloud} < 0.6$ the broadening from the low column-density
wing dominates so that the noisy PDF is broader than the underlying
cloud PDF. For wider cloud distributions, the steepening at
large column densities, visible in Fig.~\ref{fig_noise_on_full_PDF},
dominates so that the measured PDF width becomes narrower than the
cloud PDF.} At large noise
amplitudes, the measured width becomes almost independent of the
input cloud properties, saturating at $\sigma\sub{log-normal}
= 0.5 \dots 0.7$. 

Figure~\ref{fig_fullnoisescan} contains all information that is needed
to deduce the cloud properties $N\sub{peak}$ and $\sigma\sub{cloud}$
from a measured noisy PDF when knowing the noise amplitude. For any
value of $\sigma\sub{noise}/N\sub{peak}$ in the plot, we can transform
the measured peak column density $N\sub{center}$ into the corresponding
logarithmic parameter $\eta\sub{center} = \ln(N\sub{center}/\sigma\sub{noise} \times \sigma\sub{noise}/N\sub{peak})$ and compare it with the
values given in the the upper plot in Fig.~\ref{fig_fullnoisescan}.
The measured width $\sigma\sub{log-normal}$ can be directly looked up
in the lower plot. Hence, we can deduce the input parameters 
characterizing the cloud PDF $\sigma\sub{noise}/N\sub{peak}$ 
and $\sigma\sub{cloud}$ from a fit
to the measured parameters $\sigma\sub{noise}/N\sub{center}$
and $\sigma\sub{log-normal}$. The quality of the fit can be
quantified in terms of the quadratic deviations of both parameters
\begin{eqnarray}
\chi^2 & = & \left(\frac{\sigma\sub{noise}}{N\sub{center,meas}}
-\frac{\sigma\sub{noise}}{N\sub{center,Fig.5}} \right)^2  \nonumber \\
&& + \left( \sigma\sub{log-normal,meas} - \sigma\sub{log-normal,Fig.5}
\right)^2
\label{eq_chisquaredefinition}
\end{eqnarray}
where we assume the same relative accuracy for both parameters. 

   \begin{figure}
   \includegraphics[angle=90,width=\columnwidth]{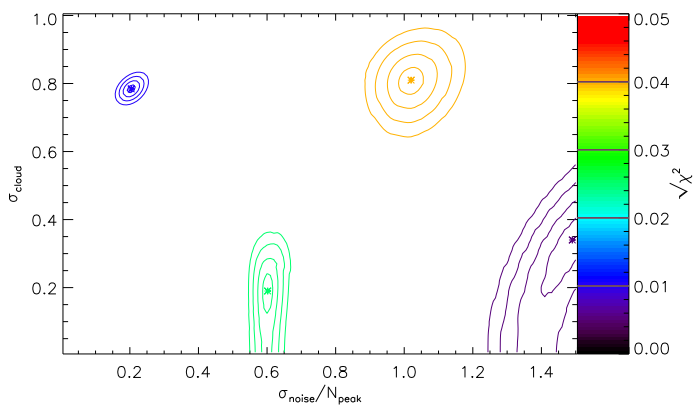}
      \caption{$\sqrt{\chi^2}$ surfaces for the fit of the 
      measurable parameters
	$\sigma\sub{noise}/N\sub{center}$ and $\sigma\sub{log-normal}$
	through the cloud parameters 
	$\sigma\sub{noise}/N\sub{peak}$ and $\sigma_{\eta,{\rm cloud}}$
	for 4 different noise-contaminated clouds with log-normal 
    column-density PDFs. Blue contours stand for a cloud
    with $\sigma\sub{cloud}=0.8$ and $\sigma\sub{noise}=0.2 N\sub{peak}$;
    green contours for $\sigma\sub{cloud}=0.2$, 
    $\sigma\sub{noise}=0.6 N\sub{peak}$; yellow contours for
    $\sigma\sub{cloud}=0.8$, $\sigma\sub{noise}=1.0 N\sub{peak}$;
    and purple contours for $\sigma\sub{cloud}=0.2$,
    $\sigma\sub{noise}=1.4 N\sub{peak}$. They show the square root of 
    the sum of the quadratic deviations of both observables
	from the measured values (see Eq.~\ref{eq_chisquaredefinition})
    at levels $\sqrt{\chi^2}= 0.01, 0.02, 0.03$, and 0.04.The
    asterisks mark the minima.}
         \label{fig_fitparam}
   \end{figure}

For any measured pair of $\sigma\sub{noise}/N\sub{center,meas}$ and
$\sigma\sub{log-normal,meas}$ we obtain the cloud parameters from
the location of the $\chi^2$ minimum in the $\sigma\sub{noise}/N\sub{peak}$ 
-- $\sigma\sub{cloud}$ parameter space\footnote{At {\tt
http://www.astro.uni-koeln.de/ftpspace/ossk/ noisecorrectpdf} we provide
an IDL program that uses the precomputed surfaces from
Fig.~\ref{fig_fullnoisescan} to provide such a fit for any input map
and given noise rms. There the Gaussian fit is only performed down
to $1/2$ of the peak to minimize the effect of distortions by PDF tails
and wings.}. This is visualized
in Fig.~\ref{fig_fitparam}. To save space, we combined the
$\sqrt{\chi^2}$ plots for four different models in one figure.
To cover a wide parameter range, we selected two log-normal 
cloud distributions with $\sigma\sub{cloud}=0.8$ and 0.2
and contaminated the first one with noise levels of $0.2 N\sub{peak}$
(blue) and $1.0 N\sub{peak}$ (yellow) and the latter one with noise 
of $0.6 N\sub{peak}$ (green) and $1.4 N\sub{peak}$ (red). The
contours show levels of $\sqrt{\chi^2}$ surfaces from 0.01 to 0.04. 

As we fit the observational parameters $\sigma\sub{noise}/N\sub{center,meas}$ 
and $\sigma\sub{log-normal,meas}$ in the space of the cloud parameters 
$\sigma\sub{noise}/N\sub{peak}$ and $\sigma\sub{cloud}$, having the same
units, we can directly read the accuracy of the parameter determination
from the topology of the $\sqrt{\chi^2}$ surfaces. For the example of
a low noise contamination ($\sigma\sub{noise} = 0.2~N\sub{peak}$)
and broad cloud distributions ($\sigma_{\eta,{\rm cloud}} = 0.8$),
the contours are quite round and the $\sqrt{\chi^2}=0.04$ contour has 
a diameter of about 0.08 in terms of the cloud parameters showing
that the cloud parameters are recovered with almost the same accuracy 
to which the observational parameters could be measured. If we increase
the noise contamination instead to $\sigma\sub{noise} = 1.0~N\sub{peak}$
the accuracy of the parameter recovery drops by a factor of three, seen
by the three times bigger diameter of the contours. For the model with
the narrow cloud distribution, $\sigma_{\eta,{\rm cloud}} = 0.2$, the
parameter retrieval is limited by the degeneracy of the measured cloud 
width, $\sigma\sub{log-normal}$, seen as extended
green plateau in the lower right corner of Fig.~\ref{fig_fullnoisescan}.
For noise amplitudes $\sigma\sub{noise} \ga 0.4~N\sub{peak}$ the
solution in terms of the cloud widths, $\sigma_{\eta,{\rm cloud}}$,
spans a very wide range while the peak position is still
well constrained. Unfortunately, this applies to many interstellar 
clouds having narrow widths $0.19 \le \sigma_{\eta,{\rm cloud}} 
\le 0.53$ \citep{BerkhuijsenFletcher2008,Hughes2013,Schneider2015a}\footnote{
\citet{BerkhuijsenFletcher2008} provided PDFs of volume densities.
The corresponding column density PDF widths can be derived following
\citet{BruntMacLow2004} and \citet{Brunt2010a}}.
In contrast, for values of $\sigma_{\eta,{\rm cloud}}>0.5$ we rather 
find a gradual decrease of the accuracy of the fit of both parameters.

   \begin{figure}
   \includegraphics[angle=90,width=\columnwidth]{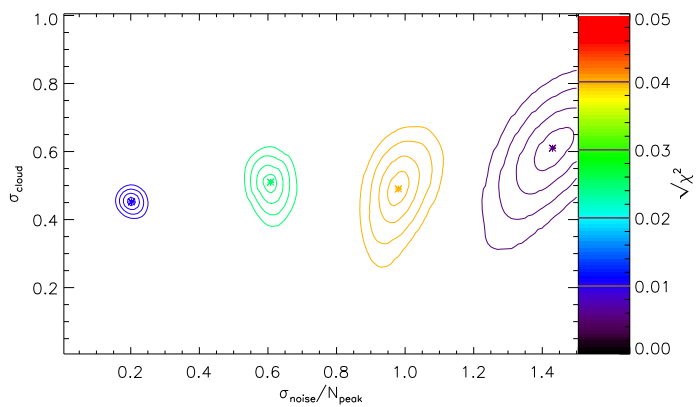}
      \caption{$\sqrt{\chi^2}$ surfaces for the fit of the measurable
      parameters
	$\sigma\sub{noise}/N\sub{center}$ and $\sigma\sub{log-normal}$
	through the cloud parameters 
	$\sigma\sub{noise}/N\sub{peak}$ and $\sigma_{\eta,{\rm cloud}}$
	for the example cloud from Fig.~\ref{fig_fractalmap} contaminated 
    with noise levels of 0.2, 0.6, 1.0, and 1.4 $N\sub{peak}$
    (blue, green, yellow, and purple contours). The contours are
    drawn at $\sqrt{\chi^2}= 0.01, 0.02, 0.03$, and 0.04. The
    asterisks mark the minima.}
         \label{fig_fitmodel}
   \end{figure}

Figure~\ref{fig_fitmodel} shows the result 
for our example cloud from Fig.~\ref{fig_fractalmap} with 
$\sigma_{\eta,{\rm cloud}}=0.45$ and the additional
power-law tail. We show again four different fits in 
one figure providing the $\sqrt{\chi^2}$ surfaces for models
with a noise contamination amplitude of 0.2, 0.6, 1.0, and 
1.4~$N\sub{peak}$, shown as blue, green, yellow, and red contours, 
respectively.

Here, we are at the edge of the parameter range dominated
by the cloud width degeneracy. For a low noise contamination
($\sigma\sub{noise} < 0.3~N\sub{peak}$) the input parameters 
are accurately recovered. At a noise level of $\sigma\sub{noise}
= 0.6~N\sub{peak}$ the accuracy of the parameter determination drops
by about a factor of two, but at higher noise levels, the
cloud-width degeneracy of the observed PDF also produces an
elongated parameter space for the solution, strongly reducing the
accuracy of the recovery of the actual cloud PDF width,
$\sigma\sub{cloud}$.

For noise contamination levels $\sigma\sub{noise}>N\sub{peak}$
the determination of the cloud parameters becomes very uncertain.
An uncertainty of 0.05 in the PDF center position of log-normal
width translates into a ten times larger uncertainty of the
fitted cloud parameters.
Altogether, we can reliably determine $N\sub{peak}$ from the map and a given noise
amplitude if the noise rms is smaller than the PDF
peak column density; for a reliable determination of the PDF width,
the noise amplitude has to fall below about $0.4 \times N\sub{peak}$.

\subsection{The zero-column-density PDF}
\label{sect_zero_pdf}

The situation is different if the noise level of the observations is 
a priori not known, e.g. in continuum data with unknown receiver
sensitivity.
In this case, the PDF may first be used to estimate the noise
in the observational data, using the probability of zero 
intensities/column densities. As {\changed any} log-normal PDF (Eq.~\ref{eq_lognormal})
has a zero probability of zero column densities, due to
the scaling with $N$ (Eq.~\ref{eq_peta_definition}), one can
attribute all measured values of zero column density to the
observational noise. Thus we can measure the noise amplitude
through inspecting the linear-scale PDF, $p_N(0)$.
Establishing the relation between the noise amplitude
$\sigma\sub{noise}/N\sub{peak}$ and the zero-column PDF,
$p_N(0)$, is also required for using $p_N(0)$
in a second step to measure the line-of-sight contamination
in Sect.~\ref{sect_contamination_simulation}.

   \begin{figure}
   \includegraphics[angle=90,width=\columnwidth]{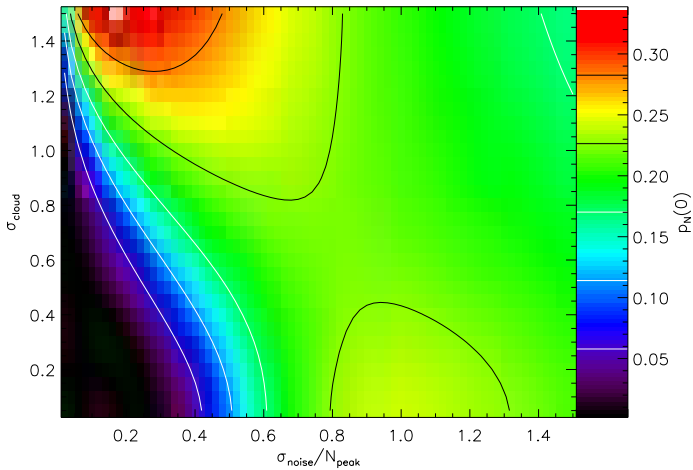}
      \caption{Values of the linear column density PDF $p_N$ at
	zero column densities as a function of the observational
	noise level and the cloud PDF width.
	The PDF values $p_N(0)$ are given in units of $1/N\sub{peak}$.}
         \label{fig_noise-zero-values}
   \end{figure}

Figure~\ref{fig_noise-zero-values} shows the 
value of the linear column density PDF for $N=0$ from the convolution 
of log-normal cloud PDFs {\changed with a normal noise PDF as a 
function of the noise amplitude and the cloud PDF width, $\sigma_{\eta,{\rm cloud}}$}. 
For very small noise amplitudes and narrow cloud
PDFs, $p_N(0)$ still vanishes, but there is a systematic increase 
{\changed when increasing noise amplitude or cloud PDF width.
For narrow cloud PDFs, the zero column density PDF peaks
when the noise amplitude matches the typical cloud density;
for wider cloud PDFs, the peak is shifted to lower noise amplitudes.
Once the combination of noise amplitude and cloud width exceeds
some threshold, we find only a small residual variation of the
zero-column PDF between values by a factor of less than two.

Closer inspection shows that the noise dependence of $p_N(0)$ can be
approximated by another log-normal function. In Appx.~\ref{appx_zero_pdf}
we demonstrate how this approximation can be used to express the 
surface in Fig.~\ref{fig_noise-zero-values} in terms of the $\sigma_{\eta,{\rm cloud}}$
dependence of three parameters only.

Instead of using Fig.~\ref{fig_noise-zero-values} to read the
zero-column PDF as a function of noise level and cloud PDF width
we can use it inversely to look up the noise level for any 
measured $p_N(0)$ when knowing $\sigma_{\eta,{\rm cloud}}$.
From the figure it is clear that this works only reliably for
low noise levels providing a steep $p_N(0)$ dependence on
the noise amplitude. For large noise levels, $p_N(0)$ does
not vary much exhibiting even an ambiguity at the largest levels.
When using the zero-column PDF to determine $\sigma\sub{noise}$
one may need an iteration with the fitting program described in the 
previous section to correct for the deviation of the actual
cloud PDF width $\sigma_{\eta,{\rm cloud}}$ from the measured width 
$\sigma\sub{log-normal}$.}

Overall we can provide a strategy for determining the parameters
of a cloud PDF from a noisy measurement. Power-law tails do not need
any correction as they are unaffected by observational noise. If the
noise level of the observation is known, the cloud parameters can
be fitted from the measured values of $\sigma\sub{noise}/N\sub{center,meas}$
and $\sigma\sub{log-normal,meas}$ using the procedure described
in Sect.~\ref{sect_noise_correction}. If the noise of the observations 
is not known, it can be determined first from the zero-column 
PDF $p_N(0)$ described by log-normal distributions having the parameters
given in Fig.~\ref{fig_noise-zero-fit}.

\subsection{Deconvolution}

   \begin{figure}
   \includegraphics[angle=90,width=\columnwidth]{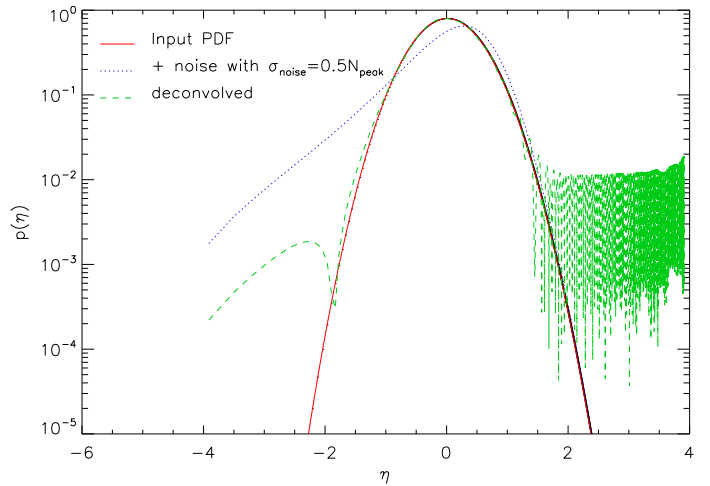}
      \caption{Result of the numerical deconvolution of a log-normal PDF
	contaminated by Gaussian noise with $0.5N\sub{peak}$ amplitude.
	The red solid line shows the input PDF before the noise
	contamination, the blue dotted line the contaminated PDFs, and the
    green dashed line the result of the deconvolution. The deconvolution
    artifacts at large column densities can be ignored as the measured
    noisy PDF already reflects the cloud PDF there.}
         \label{fig_analytic_deconvolve_example}
   \end{figure}
   
   \begin{figure}
   \includegraphics[angle=90,width=\columnwidth]{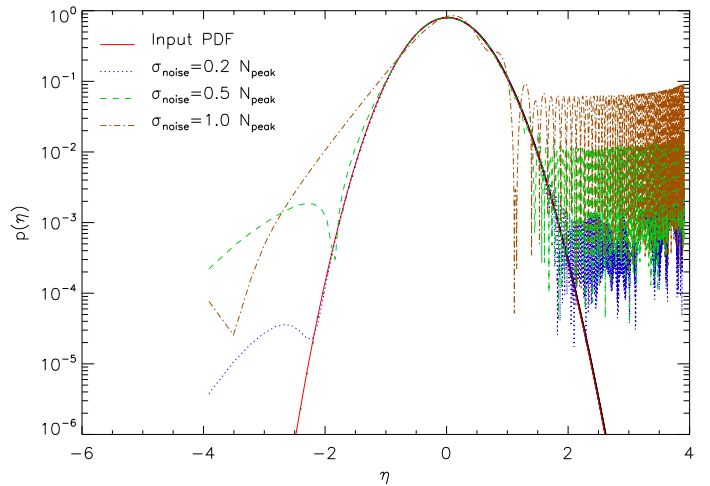}
      \caption{Result of the numerical deconvolution of a
	noise-contaminated log-normal PDF for different noise levels.
	The red solid line shows the input PDF before the noise
	contamination. The contaminated PDFs are not shown here,
	but their shape can be taken from Fig.~\ref{fig_pdfscan_noise}.}
         \label{fig_analytic_deconvolve}
   \end{figure}

As discussed in Sect.~\ref{sect_contamination} it is in principle also
possible to recover the full original PDF from the measured PDF when
knowing the noise contamination distribution.
This would also allow to recover the shape of more complex PDFs
than the simple log-normal PDF (with power-law tail) used in the 
computations above\footnote{\citet{Russeil2013} and \citet{Schneider2015b}
show some PDFs with small double-humps.}. 
Gaussian noise is the ideal case with an analytic distribution
that is well confined in Fourier space allowing to minimize the 
fundamental instability of the deconvolution process.
Figure~\ref{fig_analytic_deconvolve_example} shows one example for
the numerical deconvolution of a noise-convolved PDF for a cloud
with $\sigma_{\eta,{\rm cloud}}=0.5$ and a noise amplitude of $\sigma\sub{noise}=
0.5N\sub{peak}$ sampled in bins of $0.01 N\sub{peak}$.
As discussed before, the noise contamination 
shifts the peak of the distribution and creates a wide low column-density
tail, but leaves the high column-density part unchanged. The
deconvolution through Eq.~(\ref{eq_deconvolution}) perfectly
recovers the central part of the PDF, but creates artifacts
at low and high column densities. They result from numerical noise
in sampling the wings of the Gaussian in the denominator of
Eq.~\ref{eq_deconvolution}, i.e. the fundamental instability
in the the division of two very small numbers.
To see how the artifacts change with
contamination we show the result of the numerical deconvolution of 
noise-convolved PDFs for three different noise amplitudes in
Fig.~\ref{fig_analytic_deconvolve} (see Fig.~\ref{fig_pdfscan_noise}
for the noise-contaminated PDFs).

As the deconvolution has to be performed on a linear scale, the
approach always suffers from a low sampling at small column densities.
There the original PDF is well recovered for noise
amplitudes smaller than the peak column density. We even 
have a dynamic range of almost two
orders of magnitude over which the PDF can be reliably measured.
Only when increasing 
the noise amplitude to the PDF peak column density, the
deconvolution results in a too broad PDF after the deconvolution. 
In all cases there is, however, a floor of deconvolution artifact 
values at high column densities. Practically, those  
artifacts at large column densities are not relevant as we can
always return to the measured PDF where the noise
contamination only changed the PDF at low column densities.
At large column densities the measured PDF still reflects the 
true cloud PDF (see Fig.~\ref{fig_noise_on_full_PDF}).

   \begin{figure}
   \includegraphics[angle=90,width=\columnwidth]{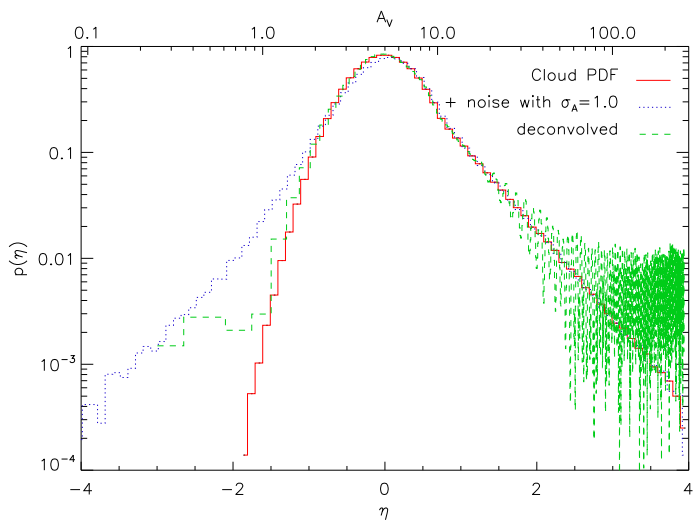}
      \caption{Result of the numerical deconvolution of the 
	cloud PDF from Fig.~\ref{fig_noise_on_full_PDF} for a noise level
	of $A\sub{V}=1$, i.e. 0.2~$N\sub{peak}$.
	The red solid line shows the input PDF before the noise
	contamination.}
         \label{fig_fbm_deconvolve}
   \end{figure}

As a more practical example, Fig.~\ref{fig_fbm_deconvolve} shows the
result of the numerical deconvolution for the full PDF from the map 
in Fig.~\ref{fig_fractalmap}. The noise amplitude of $A\sub{V}$=1
corresponds to $0.2 N\sub{peak}$, i.e. the blue dotted curve in
Fig.~\ref{fig_analytic_deconvolve}. We find a very similar behavior 
to the semi-analytic case. The dynamic range of the recovered PDF 
is a factor ten smaller than in the case of the ideal log-normal
cloud representation at the same noise level, but still a factor 
of almost 100. 

The direct deconvolution needs an accurate knowledge of
noise distribution, i.e. works only for perfectly Gaussian noise. If
the noise properties are well known, the approach may allow us to
recover small distortions and deviations of the PDF from the log-normal
shape, but to recover the four main parameters for a log-normal
PDF with power-law tail the iterative approach described in
Sect.~\ref{sect_noise_correction} is sufficient and more stable.

\section{Line-of-sight contamination}
\label{sect_contamination_simulation}

\citet{Schneider2015a} simulated the effect of a 
foreground contamination of a cloud PDF by a constant ``screen''.
In a more realistic scenario, however, the contaminating structure
also has turbulent properties leading to a similar PDF as
that of the cloud to be observed. We assume that the
contaminating cloud is also characterized by a log-normal distribution.
This excludes star-forming clouds with a significant gravitationally
dominated power-law tail but should represent the typical case where
the contaminating structure is more transparent than the 
main cloud of interest, i.e. the line-of-sight emission is dominated
by the studied cloud. The resulting PDF can be 
computed from the same convolution integral (Eq.~\ref{eq_convolution}) 
as used in the previous section\footnote{Correcting for the 
distortions introduced by material with other PDFs, e.g. 
a constant gradient can be done in an equivalent way. 
It requires the convolution with a different shape of the 
contaminating PDF, $p\sub{contam}(N)$, in Eq. \ref{eq_convolution}, 
repeating our quantitative analysis for the new convolution integral.}. 

   \begin{figure}
   \includegraphics[angle=90,width=\columnwidth]{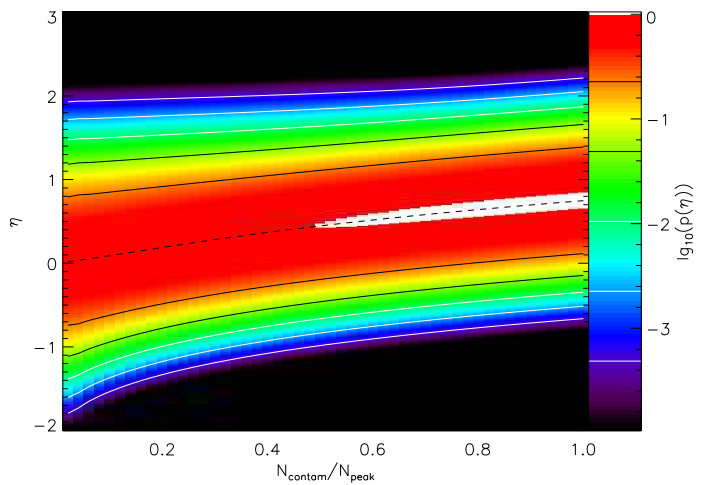}
      \caption{Distributions obtained from the convolution of a log-normal
	cloud PDF (Eq.~\ref{eq_lognormal}, $\sigma_{\eta,{\rm cloud}}=0.5$) 
	with a second log-normal PDF (Eq.~\ref{eq_contamin}, 
	$\sigma\sub{contam}=0.5$) as a function of the typical density
	of the contaminating structure, $N\sub{contam}$, relative to the peak of the
	log-normal cloud distribution, $N\sub{peak}$. The left edge of the
	figure represents the original cloud PDF shown in colors
	of the logarithm of $p_\eta$. When increasing the contamination level
	$N\sub{contam}$ in the right direction, we see the shift of the
	PDF towards higher logarithmic column densities, $\eta$.}
        \label{fig_pdfscan_contamin}
   \end{figure}

Figure ~\ref{fig_pdfscan_contamin} shows the distribution of PDFs 
obtained from the convolution when using log-normal distributions 
for the cloud PDF and the contamination as a function of the 
contamination level. In this example
both distributions have the same width of the log-normal PDF $\sigma=0.5$.
The main effect is the same as from the contamination with a constant
foreground. With increasing contamination level, the peak of the PDF
is shifted towards higher column densities and the width of the 
PDF becomes narrower. {\changed In contrast to some statements 
in the literature, line-of-sight contamination of multiple log-normal PDFs
does not create multiple peaks, but the convolution integral only
creates a broader distribution on the linear scale.} One should note that
the systematic effect on the PDF peak has the same direction for noise 
and foreground contamination, i.e., both shift the peak artificially to 
higher column densities, but that the opposite {\changed can be true for 
the standard deviation of the PDF on the logarithmic scale, where
noise increases the measured standard deviation for all narrow cloud
distributions}, while foreground contamination tends to decrease it.
The shift of the PDF peak by the line-of-sight contamination is 
approximately proportional to the contamination level $N\sub{contam}$.

   \begin{figure}
   \includegraphics[angle=90,width=\columnwidth]{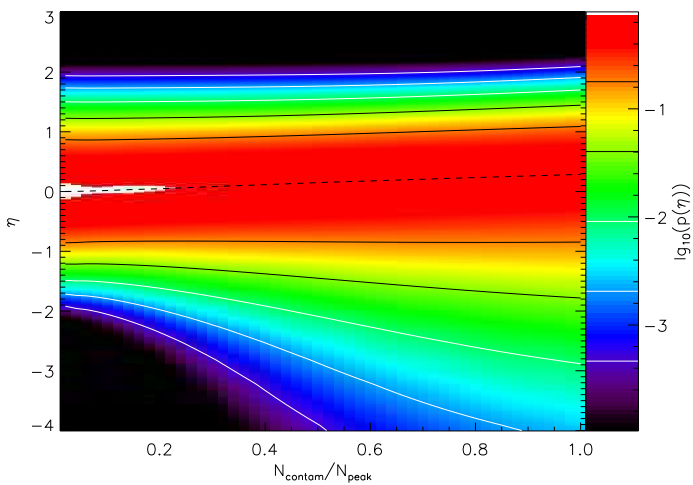}
      \caption{PDFs of contaminated clouds corrected for the contamination
	by the subtraction of a constant offset of the peak of
	the contaminating structure PDF. The PDFs are plotted as a function 
	of the density of the contamination, $N\sub{contam}$, using
	a fixed width of the contamination PDF of $\sigma\sub{contam}=0.5$
	matching the width of the main cloud PDF. The corrected distributions are
	represented through colors showing the logarithm of the PDF.} 
         \label{fig_pdfscan_contam_subtracted}
   \end{figure}

This suggests that a correction by a constant offset, as applied for the 
foreground ``screen'' by \citet{Schneider2015a}, may also work for the 
contamination
with a wider distribution. Figure~\ref{fig_pdfscan_contam_subtracted}
shows the result after subtracting the column density of the peak
of the log-normal contamination. Overall we find a quite good
reproduction of the central part of the original PDF even for
contaminations that have the same amplitude as the cloud structure
itself. The PDF peak position is recovered within $\Delta \eta =0.2$.
There is, however, a residual broadening of the distribution, in
particular towards lower column densities, due to the ``overcompensation''
of contamination contributions with lower than typical column densities.
The low column density wing of the corrected PDF appears  
too shallow relative to the original cloud PDF. One has to take into
account, however, that the plot is given in logarithmic units, i.e.
the deviations occur at levels of less than 1\,\% of the PDF peak.

   \begin{figure}
   \includegraphics[angle=90,width=\columnwidth]{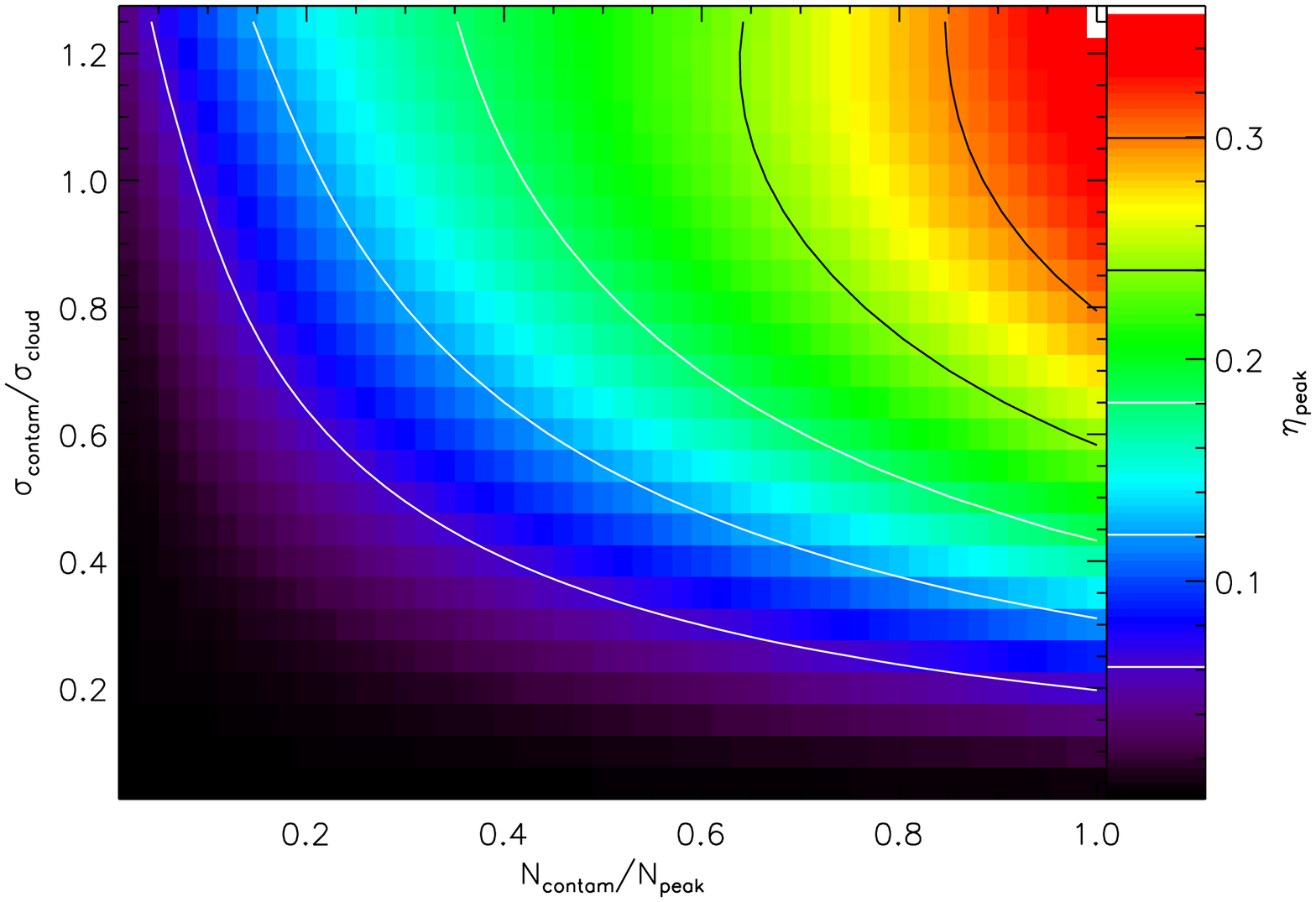}
	\vspace{0.3cm}\\
   \includegraphics[angle=90,width=\columnwidth]{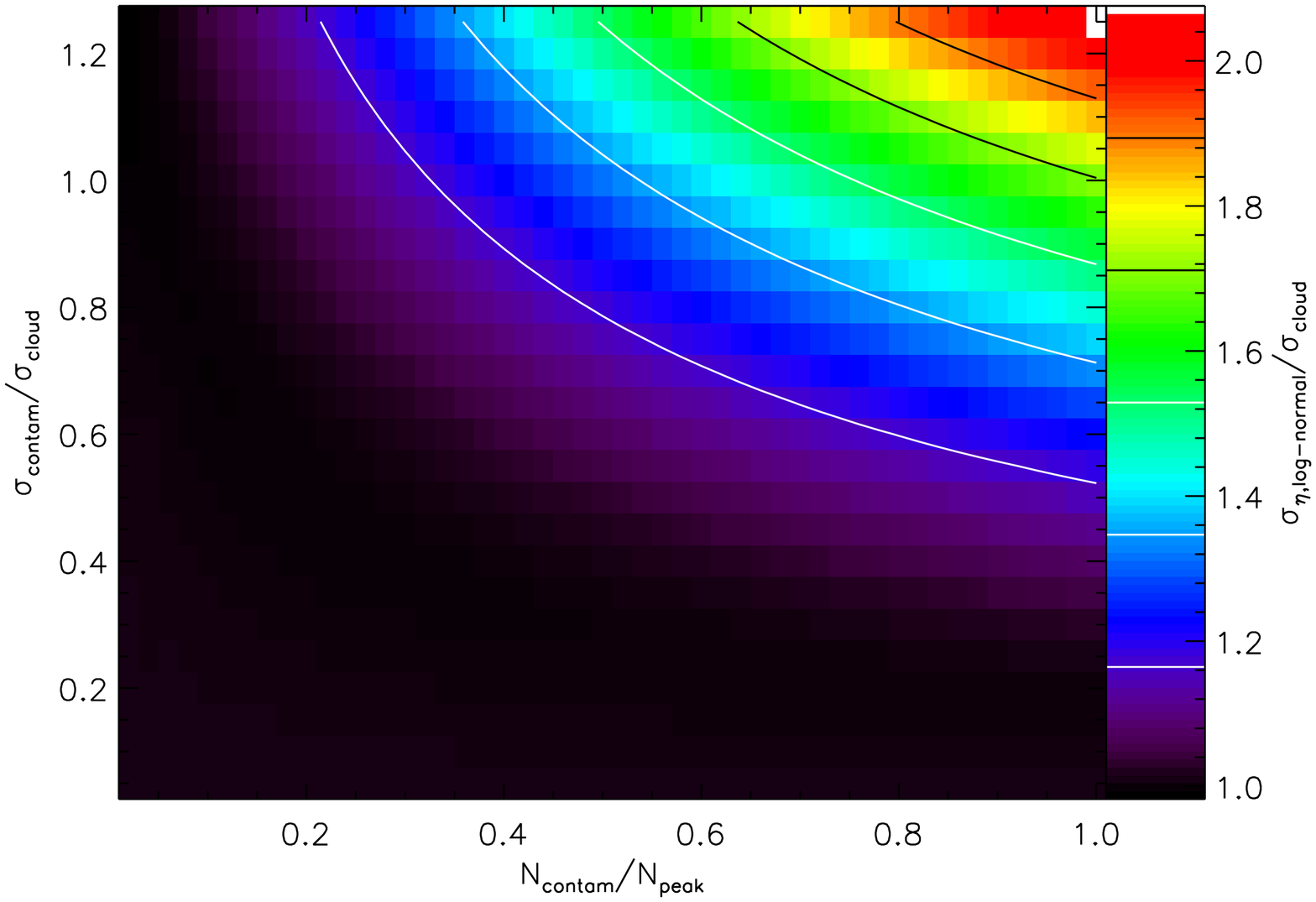}
      \caption{Parameters of the PDFs of contaminated clouds corrected 
	for the contamination through the subtraction of a constant offset
	given by the peak of the PDF of the contaminating structure. 
    The upper plot shows the
	position of the PDF peak on the logarithmic $\eta$ scale, i.e.
	a value of 0 represents the correct peak position and a value of 0.3
	stands for a 35\,\% overestimate of the peak density. The lower
	plot shows the width of the corrected cloud PDF relative to the
	original cloud PDF. In both plots we varied the amplitude of the
	contamination in horizontal direction and the width of the
	contaminating PDF in vertical direction. } 
         \label{fig_contam_subtracted_params}
   \end{figure}

To systematically quantify the residual effects after the correction,
we run a parameter scan over the full range of free parameters
covering the relative amplitude of the contamination relative
to the cloud investigated, i.e. the relative distance of the two
log-normals on the $\eta$ scale, and the relative width of the 
PDF of the contaminating structure compared to the one of the 
cloud studied. This provides two dimensional plots that are shown 
in Fig.~\ref{fig_contam_subtracted_params}.
The upper plot quantifies the accuracy of the recovery of the PDF 
peak, the lower plot shows the width of the corrected cloud PDF 
relative to the original cloud PDF. Both quantities are computed from
a Gaussian fit to the corrected PDFs. The figure shows that
the cloud PDF is accurately recovered when the width of the contaminating
PDF is narrow ($\sigma\sub{contamin} \la 0.5 \sigma_{\eta,{\rm cloud}}$) or its
column density is small ($N\sub{contamin} \la 0.2N\sub{peak}$). 
Fortunately, the first condition is usually met if the contamination 
stems from diffuse \HI{} clouds in the Milky Way, having typical PDF widths
$\sigma\sub{contam} \approx 0.1-0.2$ \citep[][assuming a typical 
$\sigma\sub{2D} \approx 0.2-0.4 \sigma\sub{3D}$]{BerkhuijsenFletcher2008}.
Significant
deviations occur if the properties of the contaminating structure are
similar to those of the observed cloud. Then the peak column density
is overestimated by 35\,\% and the width by a factor 1.7. If the width
of the contamination PDF gets even wider, this propagates directly
into the width of the corrected PDF, i.e. the recovered PDF can be 
more than twice as wide as the original PDF.

When knowing the properties of the contamination, we can use 
Fig.~\ref {fig_contam_subtracted_params} to correct this effect
in a parametric way as demonstrated above for the noise contamination.
The typical level of the contamination usually can be measured
by inspecting pixels at the boundaries of the cloud that are representative
of the contamination only. However, one has hardly ever the same statistics
for the contamination as for the whole map so that already the width of 
the contaminating PDF may not be known with sufficient accuracy.
One last resort to infer the width is the inspection of the
zero-column-density of the offset-corrected PDF, equivalent 
to our approach of measuring the noise in Sect.~\ref{sect_zero_pdf}. 
As both effects are usually superimposed onto each other, we 
first have to subtract the noise contribution to the linear PDF at 
zero column densities using the parameters from Fig.~\ref{fig_noise-zero-fit}
to derive the width of the contaminating PDF in a second step.

The procedure for this derivation is discussed in detail in
Appx.~\ref{appx_contamination}. It shows, however, that small errors 
in the determination of the amplitude of the contamination lead 
to large errors in the measured contamination PDF width.
Hence, it is practically very difficult to use the zero-column PDF 
after the line-of-sight correction to estimate the width of the 
PDF of the contaminating structure.
A direct measurement from a field tracing only the contamination seems
to be always preferable if such a region can be observed.

The direct deconvolution of the contamination from the measured PDF is
practically irrelevant as we can never know the properties of the
contamination accurately enough to allow for the computation of
the deconvolution integral (Eq.~\ref{eq_deconvolution}) without 
significant errors.
Altogether, we find that the correction of line-of-sight contamination
from a turbulent cloud can be easily corrected in terms of 
subtracting a constant screen if the width of the PDF of the 
contaminating cloud is at most half of the width of the cloud PDF
to be investigated or if its column density is lower by a factor
of 4-5. If both conditions are not met we can still measure the
PDF peak column density with an accuracy better than 35\,\%, but have
no good handle on the PDF width of the main cloud.

\section{Boundary effects}
\label{sect_edge_effects}

A typical limitation of interstellar cloud observations is given by 
the finite map size. From multi-wavelength observations it is known
since long that there is no absolute definition of an interstellar 
cloud boundary. Depending on tracer and observational sensitivity,
i.e. molecular lines, extinction, or dust continuum, the extent of a
cloud appears different and can just be approximated by the lowest
closed contour. For convenience - and to restrict the clouds e.g. 
to their molecular gas content - certain thresholds in extinction are
proposed, such as $A\sub{V}=1$ or 2 \citep{Lada2010, Heiderman2010}.
However, in most cases maps are anyway only
centered on prominent peaks and the total area that is mapped is
limited due to observational constrains.  These finite size maps
obviously truncate the statistics of the cloud PDF. 
\citet{Schneider2015a,Schneider2015b}
discussed the impact of incomplete sampling of the PDF
by considering a truncation of the statistics at various contour
levels. This only modifies the normalization of the PDF but has no 
effect on the measured shape. As an extreme example for this
selection effect, they showed that the PDFs of Infrared Dark Clouds
only consist of a power-law tail. \citet{Lombardi2015} showed that
the PDF width is reduced when mapping smaller areas.
In real observations, however, there is often no information
about the structure outside of the mapped area. 

In contrast such a truncation can also be intended.
Studying PDFs of selected subregions within an interstellar cloud
can be a valuable tool to differentiate between the physical
processes determining the column density structure. In this way
\citet{Schneider2012, Russeil2013, Tremblin2014} showed that in the
same cloud the PDFs of quiescent subregions have a log-normal shape, 
denser cloud parts exhibit a power-law tail at high column densities,
and compressed shells at the interface to H{\sc ii}-regions provoke a 
second PDF peak.
\citet{Sadavoy2014} and \citet{StutzKainulainen2015} linked 
different PDF slopes seen in different subregions of the Perseus and 
Orion A clouds to the content of young stellar objects and proposed 
an evolutionary sequence within the cloud.
Hence, it is important to statistically quantify the impact of 
the finite size maps on the properties of the measured cloud PDF
following the typical observer's approaches.

   \begin{figure}
   \includegraphics[angle=90,width=\columnwidth]{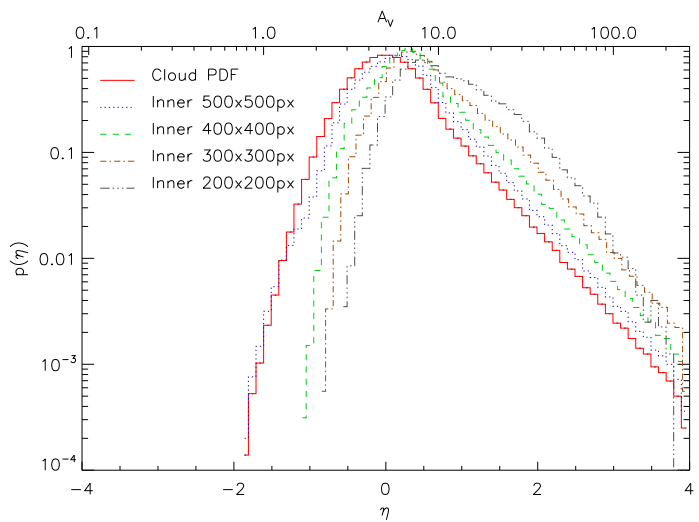}
      \caption{Result of partial observations focusing on the center
	of the cloud. PDFs are measured for the four submaps of different 
    size indicated in Fig.~\ref{fig_fractalmap}. Up to a map size 
    of $300\times 300$
	pixels, all high-column-density spots are included in the
	map. A smaller map truncates some parts of the high-density 
    center of the map leading to a noticeable distortion of the
	power-law tail. The submap sizes correspond to 69\,\%, 44\,\%,
    25\,\%, and 11\,\% of the original map area.}
         \label{fig_edge-effect}
   \end{figure}

This is simulated in Fig.~\ref{fig_edge-effect} where we study
submaps of Fig.~\ref{fig_fractalmap} with different size, placed
around the central column density peak.
In this simulation the spatial structure of the cloud starts 
to become relevant while in the previous sections the convolution 
integrals were actually independent of the cloud shape. Details
of the resulting PDFs may depend on the exact shape of the
mapped structure, but the general behavior is the same for all cases 
if we follow a typical observational approach. Our submaps try to 
mimic the observational strategy for mapping an interstellar cloud structure 
around the central density peaks with a finite array in a 
limited observing time. 

As long as a sufficiently large part of the cloud is covered,
the truncation at the boundaries only removes low column density gas
from the statistics, leaving the power-law tail unchanged -- except 
for the modified normalization. The removal of low-density pixels
also shifts the peak of the log-normal part to larger column densities.
The effect is small as long as less than half
of the pixels are removed, but already when going from
$600\times 600$ pixels to $400\times 400$ pixels (44\,\% of the area)
the truncation of the low-density statistics becomes significant so 
that we significantly overestimate the peak position. In that step
one can also notice a measurable reduction of the PDF width
that continues when going to smaller submap sizes. At a submap
size of 11\,\% of the original map, we also start to loose some
of the high-density structures. In this case, even the power-law
tail is affected.

The impact is much stronger when considering mean quantities.
This is a well know effect to most
observers: increasing the mapping size around the main
emission peaks tends to drastically lower the average
column density as more ``empty'' regions are added to the
statistics. For our example in Fig.~\ref{fig_edge-effect}
we find that the measured peak of the PDF in logarithmic bins $N\sub{center}$
changes by a factor of 1.6 when reducing the field size from 
$600 \times 600$ to $200\times 200$ pixels (11\,\% of the area), 
the same factor applies for the peak in linear units, but that the average
column density increases by a factor of 2.3 when reducing the 
map size\footnote{As our test cloud also includes a power-law
tail of high densities, the average column density is not
only 10\,\% higher than $N\sub{peak}$ as expected for
a log-normal distribution (see Sect. \ref{sect_pdf_math}), but 
44\,\% higher for the original map.}.

   \begin{figure}
   \includegraphics[angle=90,width=\columnwidth]{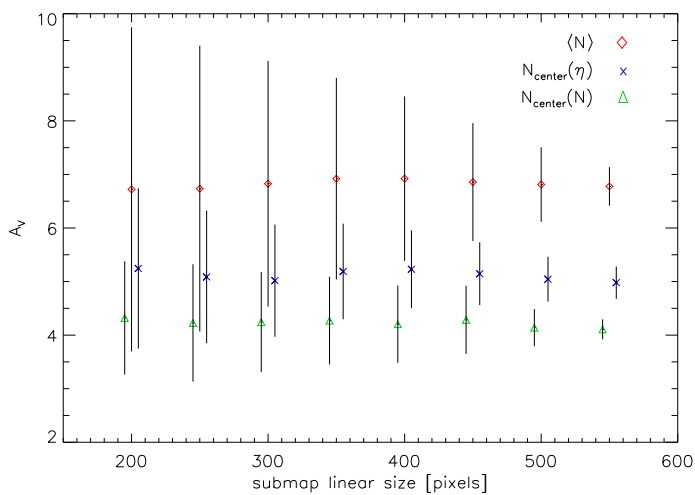}
      \caption{Parameters of the column density distribution
      	of randomly selected submaps of the map from Fig.~\ref{fig_fractalmap}
        as a function of their size. The symbols show the mean from an ensemble
        of 50 submaps and the error bars represent the standard deviation.
        Red diamonds show the average
        map density, blue crosses give the PDF peak on the logarithmic
        scale, and green triangles the peak on the linear density
        scale. The symbols and error bars for the two PDF peaks are 
        displaced by $\pm 5$ pixels relative to the actual submap size
        for a better visibility in the plot.}
         \label{fig_edge-density}
   \end{figure}
   
To statistically quantify this selection effect we investigate ensembles of
50 randomly selected submaps of different size and measure
their average and most probable column densities. Figure~\ref{fig_edge-density}
shows the ensemble mean and standard deviation for the average
column density in the submaps and the PDF peak position 
$N\sub{center}(\eta)$ on the logarithmic scale. For completeness, we 
also computed the center of a Gaussian PDF on the linear column 
density scale $N\sub{center}(N)$. In contrast to the
submap selection centered on the main peak in Fig.~\ref{fig_edge-effect},
the random selection provides for all three parameters a mean 
ensemble density that does not depend on the map size. 
The mean average density and PDF peak positions of the input map 
are recovered, but we find that 
the uncertainty of the parameters grows drastically towards smaller 
submap sizes. While the uncertainty of the PDF peak positions 
grows up to 25\,\% for $200\times 200$ pixel submaps, the uncertainty
of the average column density grows to 45\,\%. 

This bigger uncertainty of the average density is to be expected
as the average is more easily affected by low-probability outliers
and they are more easily missed or hit in a random selection than
values with high probabilities. This explains why the use of the 
PDF peaks for normalization purposes  (see Sect.~\ref{sect_pdf_math}) 
always provides more stable results than the use of the average density.

\section{Interferometric observations}
\label{sect_interferometer}

A special source of observational uncertainty results from the incomplete
sampling of the uv plane that is unavoidable in interferometric
observations. \citet{Rathborne2014} discussed the shape of the
low-column density tail of a PDF obtained from combining ALMA 
observations with Herschel data, but
they did not quantify to what degree that tail is determined by 
the observational limitations of the interferometric observations.

\subsection{ALMA}

\begin{figure}
   \includegraphics[angle=90,width=\columnwidth]{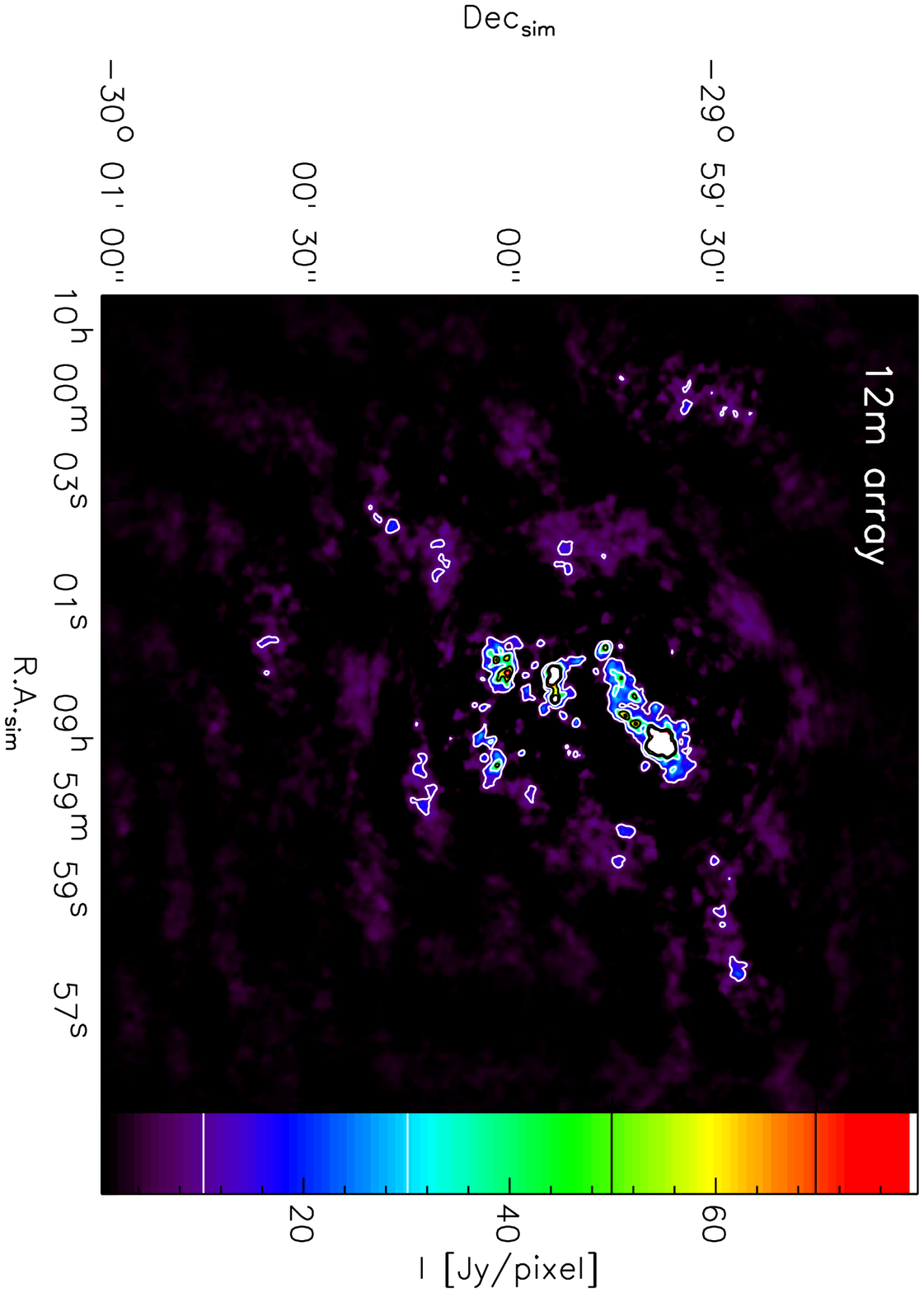}
   \vspace{0.3cm}\\ \includegraphics[angle=90,width=\columnwidth]{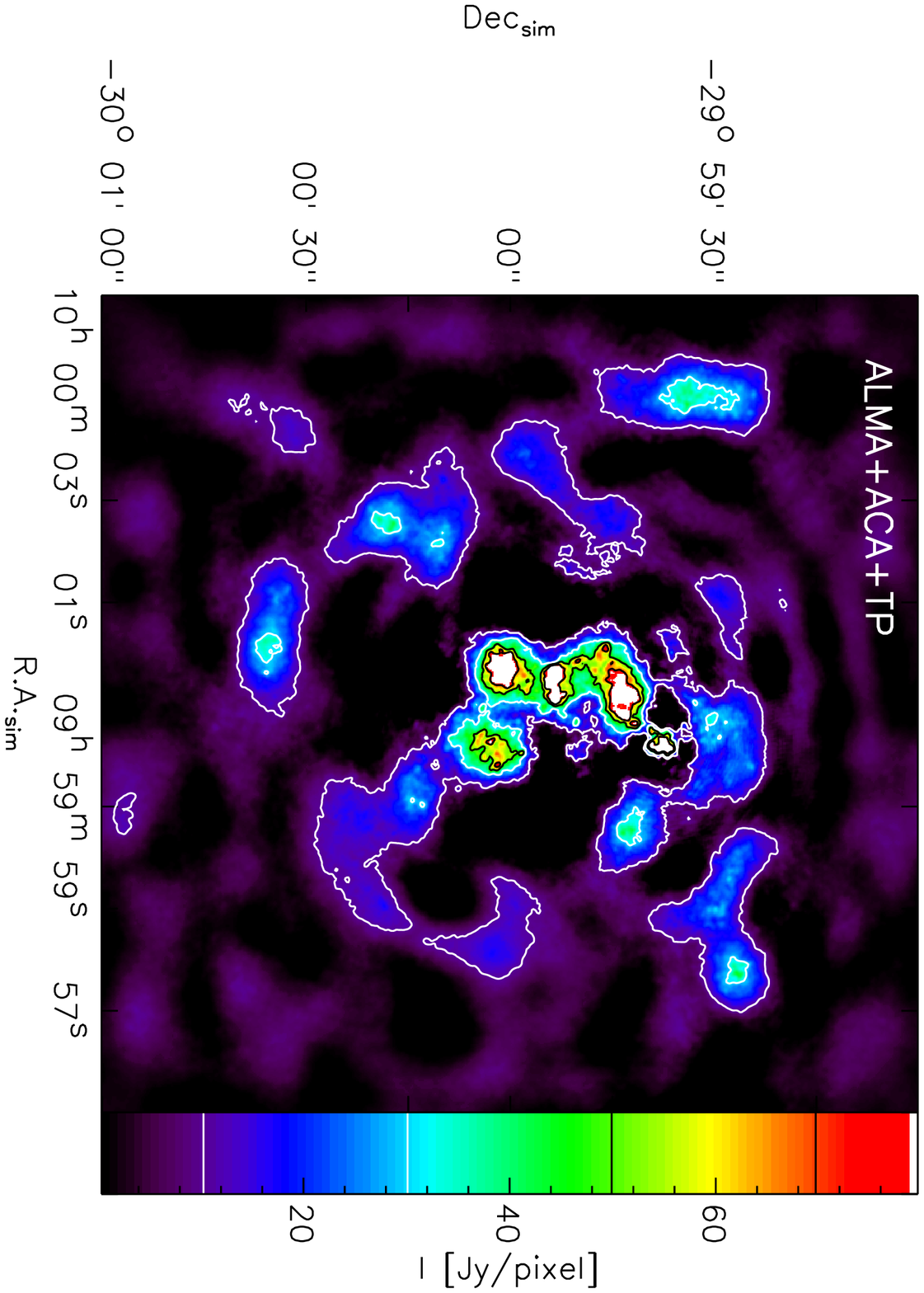}
      \caption{Result of a simulated 2h ALMA observation of the
	cloud {\changed shown in Fig.~\ref{fig_fractalmap} where 
	we translated the optical depth $A\sub{V}$ into an equivalent
	230~GHz intensity}.
	The upper plot shows the map obtained when using the 12m array
    without short-spacing correction. The lower plot is the
    map obtained after combining the 12m array data with the data 
    from the compact array and single-dish zero spacing. The 
    intermediate map obtained from combining only the ALMA 12m and 
    ACA results is visually hardly distinguishable from the map in the 
    lower plot so that it is not shown here. }
         \label{fig_ALMA-result}
\end{figure}
   
\begin{figure}
   \centering
   \includegraphics[angle=90,width=\columnwidth]{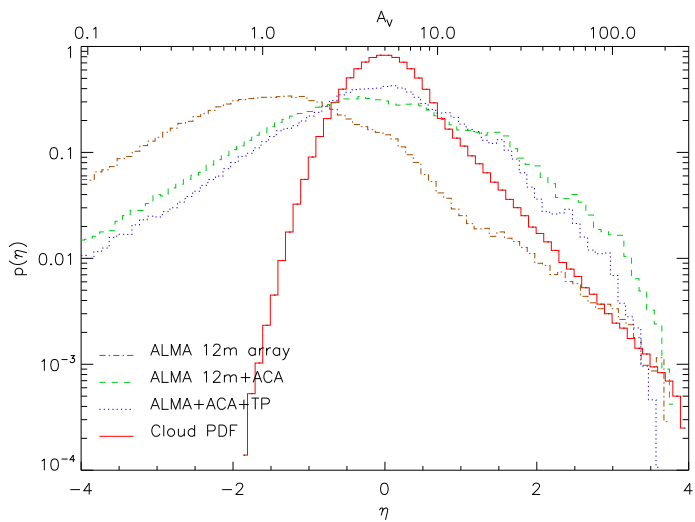}
      \caption{Comparison of the original cloud PDF with the PDFs
      of the maps obtained from the 2h ALMA observations 
      (Fig.~\ref{fig_ALMA-result}). }
         \label{fig_ALMA-pdfs}
\end{figure}

{\changed ALMA provides a better instantaneous uv coverage than
any other observatory available before, based on the unprecedentedly
large number of dishes and baselines available. However,
any incomplete information in the Fourier domain may still have 
significant effects on the measured intensity PDF.}
We simulate the behavior using
the ALMA simulator in CASA 4.4\footnote{Spot tests with CASA
3.3 and CASA 4.5 showed no significant differences.}
with a configuration providing a resolution of $0.6$''.
We used the default pipeline setup based on the standard CLEAN 
algorithm \citep{Hoegbom1974} that would be used by most
observers. To convert our column density map into observable
quantities we treated the input map as
Jy/pixel units, used 230 GHz as center frequency, and a 7.5 GHz bandwidth
corresponding to standard continuum observations. We used the mosaicking 
mode to map the total field of view of $(2')^2$, and combined the 12m 
array and ACA observations matching what ALMA provides 
for continuum observations (ALMA+ACA). The choice of these
parameters should not have any major impact on our results
as we were not limited by sensitivity or resolution as discussed 
later.
%to have reasonable intensity units usable in the simulator.
%{\bf Details to be corrected by Timea.}
To obtain a complete picture for the capabilities and limitations
of the approach, we also performed the step of the single-dish total 
power correction (ALMA+ACA+TP). Total power zero spacing is not
provided for continuum observations at ALMA, only for line
observations, but including this step provides us with a more 
complete picture in terms of the fundamental capabilities of 
interferometric observations.
In a separate test, we added 
thermal noise to the simulations but found that it does not have any
significant impact, indicating that the results are not limited by 
sensitivity but, as intended, by uv coverage.

Figure~\ref{fig_ALMA-result} shows the maps and Fig.~\ref{fig_ALMA-pdfs}
contains the corresponding PDFs from an observation that can be executed
in a 2h observing block (2h for 12m array, 4h for ACA, 8h for TP).
%We used a configuration providing a resolution of 0.6$''$ , i.e.
%three pixels and the mosaicking mode. The images were simulated at 230.GHz
%with a 7.5 GHz continuum bandwidth assuming a combination of
%12m, ACA and TP observations. 

The simulated observations with the 12m array only provide a
reasonable reproduction of the position and shape of the high-intensity
contours, but we recover basically no information about 
structures with an intensity below 20~Jy/pixel, i.e. only
the high end of the PDF power-law tail ($\eta > 3$) is retained. 
After combining the data with the short-spacing information, some 
low-intensity structure is recovered appearing in spiral shapes,
probably produced by the shape of the uv tracks. But the
shape of the high-intensity peaks is more heavily distorted than
in the pure 12m array data. The intermediate map obtained 
when combining the data from the 12m array with the ACA data
is very similar to the map obtained when including all 
short-spacing information. 
The visual impressions from the
interferometric maps are confirmed by the PDFs of the maps.
When sticking to the pure
interferometric data, the CLEAN algorithm manages to reproduce 
the high-intensity peaks above $\eta=3$ so that the 
high-intensity end of the power-law tail of the PDF is recovered. 
At lower intensities, however, there is hardly any 
correspondence of the map PDF to the underlying cloud PDF.
The typical values of $A\sub{V} \equiv 5$~Jy/pixel are strongly 
under-abundant and the PDF peak falls at a five times lower value.
{\changed All curves show an extreme excess of low intensity 
contributions, similar to the observational noise effect shown 
in Fig.~\ref{fig_noise_on_full_PDF}.
When} combining the interferometric data with the 
short-spacing information, the high-density tail is even more
distorted, but the location of the peak of the PDF is reasonably
recovered. {\changed The excess at low intensities is still high}
but five times lower compared to the 12m array data. 

\begin{figure}
   \includegraphics[angle=90,width=\columnwidth]{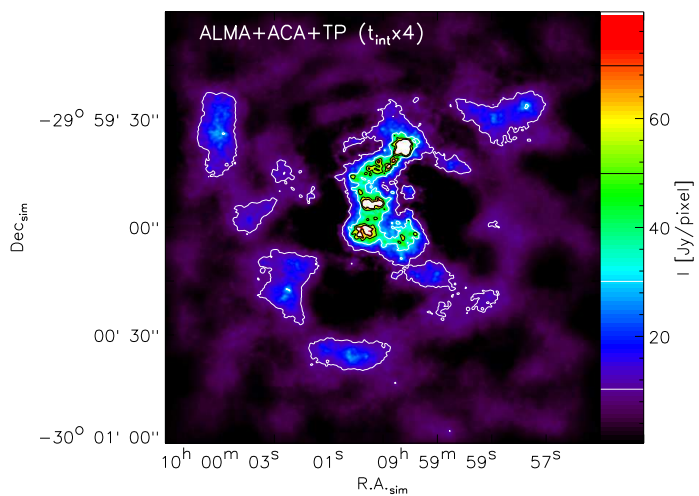}
      \caption{Result of a simulated 8h ALMA observation of the
	cloud shown in Fig.~\ref{fig_fractalmap}.
	The map shows the final product from combining the 
    12m array data with the data from the compact array and 
    single-dish zero spacing. }
         \label{fig_ALMA_long-result}
\end{figure}
   
\begin{figure}
   \centering
   \includegraphics[angle=90,width=\columnwidth]{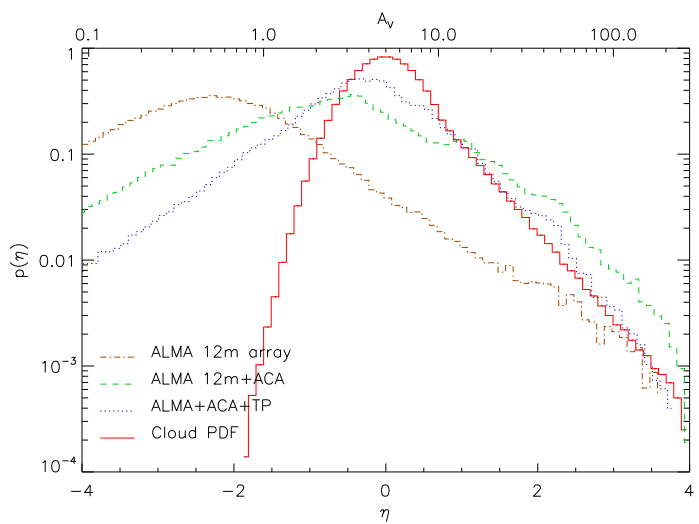}
      \caption{PDFs of the maps obtained from the 8h ALMA observations
      (Fig.~\ref{fig_ALMA_long-result}). }
         \label{fig_ALMA_long-pdfs}
\end{figure}

An obvious approach to improve the mapping results is the use of
longer integration times, automatically providing a better uv plane
coverage due to the rotation of the earth. Figures~\ref{fig_ALMA_long-result}
and \ref{fig_ALMA_long-pdfs} show the results when quadrupling 
the observing time.
Here, the simulations used a total continuous observing time of 8h 
with the 12m array. This corresponds to 40h of total observatory
time (8h for 12m array, $2\times 8$h for ACA, and $4\times 8$h for TP),
an amount that rarely would be granted. It is, moreover, an idealized
case because usually 2h observing blocks are executed randomly by the 
observatory, potentially leading to overlaps in the uv coverage.

When considering the map from the 12m array we do not find a
big improvement compared to the shorter integration time. Again only
the high intensity end ($\eta > 3$) of the cloud PDF is recovered. More
low-intensity material is detected leading to an additional shift of 
the PDF maximum to even lower intensities. Adding short-spacing 
information does no longer provide the correct PDF peak position, but
when including the single-dish zero-spacing, {\changed we fully recover
the high-intensity PDF power-law tail.} This is also visible in the 
map in Fig.~\ref{fig_ALMA_long-result} which shows a much better 
resemblance to the input map (Fig.~\ref{fig_fractalmap}), reproducing
all bright structures. The combination of a very long integration 
time with all zero-spacing information allows to actually
recover a significant part of the underlying cloud PDF. Unfortunately,
this is currently not possible in real observations as ALMA
does not provide the single-dish total power information for continuum
data yet. Moreover, we can conclude that the need for a wide uv
coverage is not well satisfied by single observing blocks of 2h.
For the cloud statistics, it would be better to distribute the same 
observing time in smaller chunks over the whole visibility 
window of a celestial object, at least if one is not limited by 
sensitivity as in our example.

Even in this case the PDF still does not contain any reliable
information on the log-normal part. It shows the same large excess
of low intensities as in the case of the shorter integration time.
The excess appears very similar to the low-intensity PDF wing reported
by \citet{Rathborne2014} for their interferometric data.
 
\begin{figure}
   \centering
   \includegraphics[angle=90,width=\columnwidth]{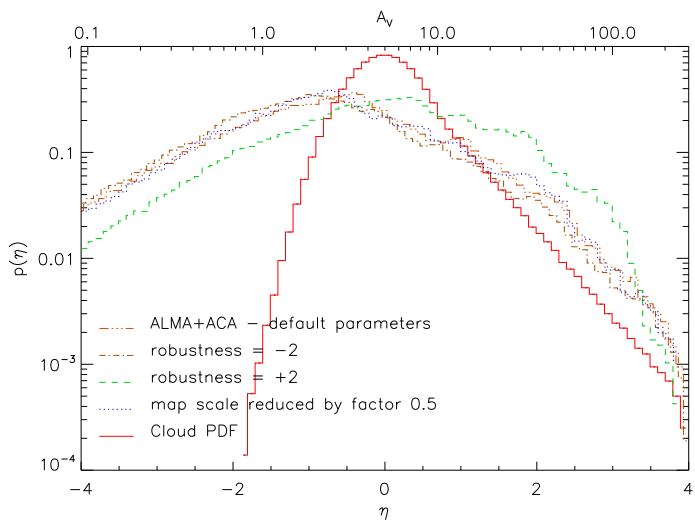}
      \caption{Impact of different parameters on the PDFs from the
      ALMA+ACA maps obtained in the simulated 8h ALMA observations.
       }
         \label{fig_ALMA-parameters}
\end{figure}

In the combination of the data from the 12m array and ACA only, the
longer integration time already leads to an improvement in the 
recovery of the high-intensity tail compared to the 2h ALMA
observation, but the deviation of the measured PDF from the cloud
PDF is still significant. One can only conclude that there must be 
a power-law tail. Measuring its parameters is still unreliable. 
To see whether one can better recover the cloud PDF based on the 
data provided by ALMA, we performed some additional 
tests varying the  
parameters of the CLEAN algorithm. The results are shown in 
Fig.~\ref{fig_ALMA-parameters}.
In two of the curves we modified the robustness parameter of the
``Briggs'' weighting (see CASA User Reference \& Cookbook\footnote{{\tt
http://casa.nrao.edu/docs/cookbook/}}). A
value of $-2$ corresponds to uniform weighting, cleaning with 
a weighting robustness of $+2$ corresponds to natural weighting. 
The figure shows that 
natural weighting clearly deteriorates the recovery of the map
PDF, there is no significant difference between the standard weighting
(robustness of 0) and the uniform weighting. In a second test we
used a mask to help the convergence of the CLEAN algorithm. This 
should provide a better data reduction strategy. However, we found
no significant difference to the outcome using the default
parameters so that the corresponding curve is not added in the
plot, being almost identical to the brown curve. Finally, we
tested the impact of the map size by changing the pixel size of
our test data set by a factor of 0.5 resulting in a map size of
$1'^2$. In this way we should quantify the pure resolution effect.
The blue curve in Fig.~\ref{fig_ALMA-parameters} shows, that 
this has also no significant impact on the recovery of the cloud
PDF. As long as we avoid the natural ``Briggs'' weighting, the 
combined ALMA+ACA maps are equally good or bad for measuring the
cloud PDF.  

A reliable partial recovery of the cloud PDF is only possible
when {\changed combining single dish information with integrations
covering a long time span to provide} a very wide uv coverage. 
This asks for an extension of the current capabilities of ALMA.

\subsection{Fundamental limitations}

{\changed
To test whether the ALMA results are due to some limitation of the
CLEAN algorithm combined with the continuous tracks given by the
telescope baselines or whether they rather reflect the general 
problem of lacking information in uv space, we added some
numerical tests with a more controlled random coverage of the
uv plane. We transformed the
cloud from Fig.~\ref{fig_fractalmap} into the Fourier domain and
randomly removed a number of the points in Fourier space. The
zero-spacing point, characterizing the mean intensity in the map, 
is always preserved.

   \begin{figure}
   \includegraphics[angle=90,width=\columnwidth]{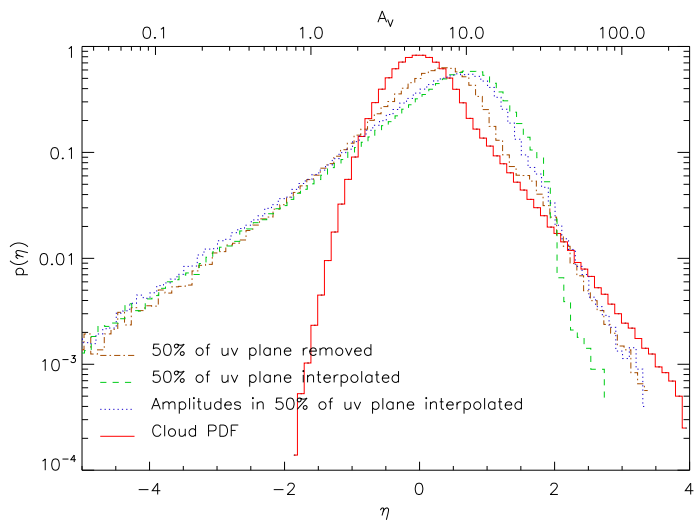}
      \caption{Result of incomplete sampling in the uv plane.
	The PDFs are obtained after removing the information for half of 
	the points randomly selected in the uv plane. We either replaced
	the Fourier coefficients by zeros, by values interpolated between 
	the retained neighboring points or by interpolated amplitudes and
	random phases.}
         \label{fig_uv-impact-interpol}
   \end{figure}

Figure~\ref{fig_uv-impact-interpol} shows the resulting PDFs after the 
inverse transform when removing the information for half of the points.
The brown dashed-dotted line is computed by replacing the missing 
information by zeros. The green dashed curve is obtained by 
interpolating the missing points in Fourier space, and the blue dotted
curve uses interpolated amplitudes but random phases. All curves show
an excess of low intensity contributions, like the ALMA maps, and a shift 
of the PDF peak to higher intensities compared to the input map. 
The highest intensity fraction is suppressed. Part of those pixels
are transformed to lower values resulting in a steepening
of the power-law tail. The worst
reproduction of the cloud PDF is obtained when interpolating in the
uv plane. The interpolation acts like a low pass-filtering that suppresses
sharp structures, therefore washing out some of the high-intensity peaks.
For a fractal cloud one expects in principle a smooth change of the
Fourier amplitudes while the phases represent the details of the
realization, but the experiment with random phases does not provide
any improvement compared to the test with the simple zero replacement. Other
tests with constant phases, systematic phase changes or conjugate complex 
numbers also did not provide any improvement. 
A recovery of the underlying cloud PDF seems to be impossible. Removing half
of the points in the uv plane has a much stronger impact on the
determination of the PDF than removing half of the points in normal
space as demonstrated in Sect.~\ref{sect_edge_effects}.
As the removal of information in Fourier space affects every single pixel
of a map, the impact on the real-space statistics is always worse compared
to removing the information for the same number of pixels in the map.

   \begin{figure}
   \includegraphics[angle=90,width=\columnwidth]{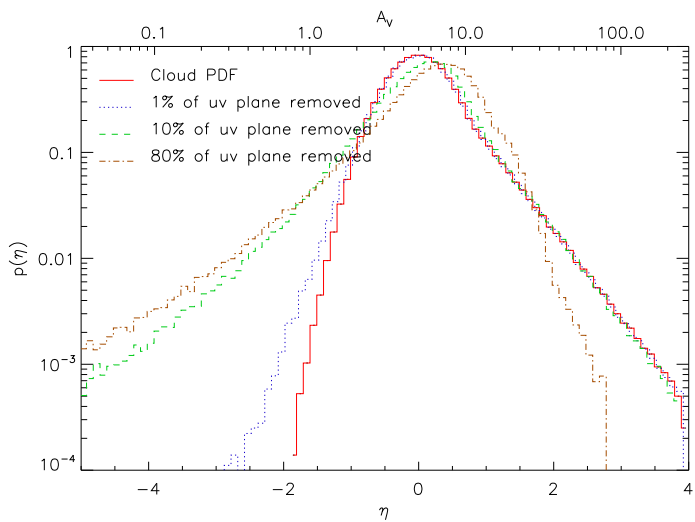}
      \caption{PDFs from a variable sampling in the uv plane. Different
	fractions of randomly selected points in the uv plane were replaced
	by zeros.}
         \label{fig_uv-coverage}
   \end{figure}

To get an idea about the minimum coverage of the uv plane needed
to obtain a reliable PDF measurement we repeated the test above,
but removed different fractions of the points in the Fourier space.
Figure~\ref{fig_uv-coverage} shows the PDFs obtained when randomly replacing
1\,\%, 10\,\%, and 80\,\% of the points by zeros. 
% Like shown for the 50\,\% removal in Fig.~\ref{fig_uv-impact-interpol}, 
% the PDFs obtained from interpolated Fourier amplitudes and random 
% phases for the missing points are very similar. 
Even when only 10\,\% of the uv plane
are not covered, we find a significant low intensity excess. In
this case PDF peak and the high-intensity tail are still well recovered.
The 80\,\% case shows hardly any similarity with the input PDF.

   \begin{figure}
   \includegraphics[angle=90,width=\columnwidth]{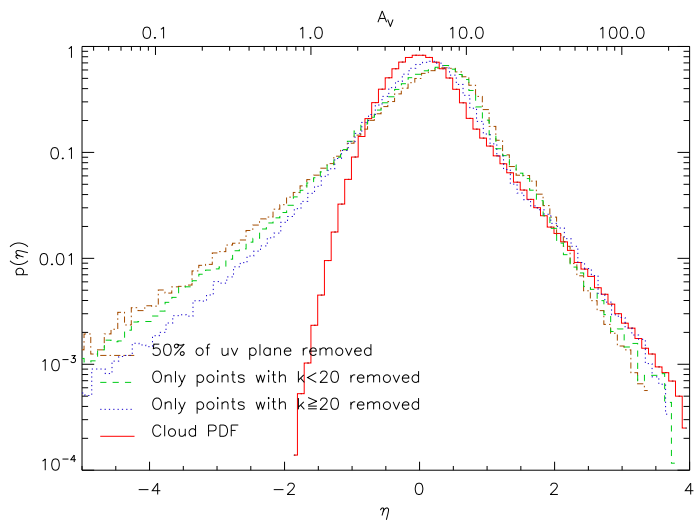}
      \caption{PDFs from an incomplete sampling of the uv plane depending
	on the wavenumber. The brown dashed-dotted line shows the result
	when randomly replacing 50\,\% of the uv plane by zeros, as shown
	already in Fig.~\ref{fig_uv-impact-interpol}. The other two PDFs are
	computed when only removing those points inside or outside of
	a circle of 20 wavenumbers, respectively.}
         \label{fig_uv-wavenumber}
   \end{figure}

To better understand the reason for this effect, we finally show the
dependence of the PDFs on the wavenumber coverage in Fig.~\ref{fig_uv-wavenumber}.
We again randomly remove 50\,\% of the points in the uv plane, like
in Fig.~\ref{fig_uv-impact-interpol}, but also restrict the removal to
the areas of small and large wavenumbers, $k \lessgtr 20$, respectively.
Due to the small area of $k<20$ in the uv plane, only 0.15\,\% of the
points are removed in this case, while for $k\geq 20$, 49.85\,\% of the
points are removed. We see that the main distortion of the PDF power
law tail and the PDF peak stems from the missing information at small
wave numbers. If we have a complete coverage of the wave numbers 
$k < 20$ we recover the PDF over a dynamic range also covering a factor of 20.
Lack of short-spacing information has a much worse effect on the PDF
than lack of high frequency information. Any incomplete uv coverage, 
however, creates a significant excess of low-intensity noise. 

Combining the picture with the information from Fig.~\ref{fig_uv-impact-interpol}
suggests that it is almost impossible to obtain any reliable information
on the low-intensity statistics from interferometric observations.
To get a measurement of the high intensity tail and the PDF peak, a 
good coverage of the low wavenumber, i.e. short-spacing range of the
uv plane is needed. The similarity in the 
PDF recovery shown in Fig.~\ref{fig_ALMA_long-pdfs} for the long ALMA
observations and in Fig.~\ref{fig_uv-wavenumber} suggests that there
is no fundamental difference between the map reconstruction by the
CLEAN algorithm and by the direct Fourier transform. The critical point
is a sufficiently dense sampling of the uv plane in particular for
low wave numbers. Only with that dense sampling one can recover 
the upper part of the PDF.
However, any sparse sampling of the uv plane unavoidably leads to 
severe distortions of the overall PDF shape.}
None of our experiments simulating interferometric observations
allowed to recover the low-intensity part of the PDF including
the log-normal distribution. Therefore, the interpretation of 
PDFs from interferometric data should be done with extreme care.
%__________________________________________________________________

\section{Conclusions}

When measuring interstellar cloud column density PDFs, the
observational effects have to be carefully treated. Otherwise,
all conclusions on the PDF parameters can be wrong by large
factors. We provide tools to derive the cloud parameters from 
the measured PDFs. Our focus lies on on the log-normal 
part of the PDF because this part is most strongly affected by 
observational noise, contamination 
by fore- or background material, and the selection of the map boundaries.
In single-dish observations the power-law tail in the PDF of self-gravitating 
clouds is typically only marginally influenced by the effects discussed
here \citep[see also][]{Lombardi2015}. This is different for
interferometric observations.

Noise can be treated mathematically as a convolution integral. 
If the noise distribution is perfectly Gaussian the original 
PDF can be recovered through direct deconvolution. 
For general single-dish data, the PDF parameters 
can be computed from the measured PDF and a known noise level
using the algorithm discussed in Sect.~\ref{sect_noise}. This 
allows for an accurate measurement of all parameters of the 
log-normal part if the noise level falls below 40\,\% of the
typical column density in the map. It even works for noise 
levels up to the typical column density in the map in case of
clouds with a broad PDF ($\sigma_{\eta,{\rm cloud}} > 0.3$). For clouds
with narrower PDFs and high noise contaminations, we can only 
retrieve the position of the PDF peak, but no longer its width.
If the noise level in a map is a priori not known it can be 
deduced from the zero-column PDF.

Line-of-sight contamination {\changed does not create multiple
peaks in the column-density PDF, but a broader distribution on
the linear scale. It} can be corrected by subtracting
a constant ``screen'' with the typical column density of the
contamination from the measurement. This works well if the
PDF of the contamination is either a factor of two narrower 
or a factor of four to five weaker than that of
the cloud to be characterized. If width and strength of the
contamination are well known, one can also retrieve the
parameters of the cloud PDF through a fit to the measured
parameters, similar to the noise correction. However, in the 
general case of an unknown broad contamination with a
column density comparable to that of the cloud studied,
we find a strong ambiguity between the impacts of the
two PDF widths so that it turns out to be impossible to reliably
derive the cloud PDF width from the measurement.
We can only recover the PDF peak position of the
studied cloud with an accuracy of about 35\,\%.

The effect of a limited field of view due to map boundaries 
cannot be easily corrected as we cannot ``invent'' information that
was not measured. We found, however, that the effect is small
if the observations cover at least 50\,\% of the structure to be
measured. Smaller maps will underestimate the PDF width and
overestimate the peak position. The width is slightly more stable
against selection effects than the position. The sensitivity to
boundary selection effects clearly discourages from the use of 
mean quantities to characterize the properties of a map. The
characterization in terms of the PDF peak is more stable.

The situation is very different for interferometric observations 
with a sparse sampling in the uv plane. We found no way
to reliably measure the statistics for the majority of the pixels 
in the map forming the center and low-density wing of the PDF. 
The removal of information in the uv plane has a much stronger 
impact than the direct removal of pixels in the map.
An interferometric determination of the high-density PDF tail is 
possible, but asks for an approach that goes beyond
the current capabilities of ALMA. Combining single-dish
zero spacing with a very long integration time or with many 
small chunks of 12m array observations over a long period, that 
widely populate {\changed in particular the short-spacing
area of} the uv plane in principle allow to retrieve
the high-density wing of a cloud PDF.

\begin{acknowledgements}
      We thank the referee Mark Heyer for very constructive suggestions that
	 helped to improve the paper.
      This work was supported by the German
	\emph{Deut\-sche For\-schungs\-ge\-mein\-schaft, DFG\/} via the 
	SPP (priority programme) 1573 'Physics of the ISM'. V.O. and N.S.
	acknowledge supported by the DFG project number OS 177/2-2. 
      C.F. acknowledges funding provided by the 
	Australian Research Council's Discovery Projects
	(grants DP130102078 and DP150104329).
    RSK acknowledges support from the DFG via projects KL 1358/18-1 and KL 1358/19-2 and via the SFB 881 ``The Milky Way System'' (sub-projects B1, B2 and B8). In addition, RSK thanks for funding from the European Research Council under the European Community’s Seventh Framework Programme (FP7/2007-2013) via the ERC Advanced Grant ``STARLIGHT'' (project number 339177).
\end{acknowledgements}

%-------------------------------------------------------------------
\bibliographystyle{aa}
\bibliography{ref}

\clearpage
\appendix

\section{Parameterization of the noisy zero-column-density PDF}
\label{appx_zero_pdf}

   \begin{figure}
   \includegraphics[angle=90,width=\columnwidth]{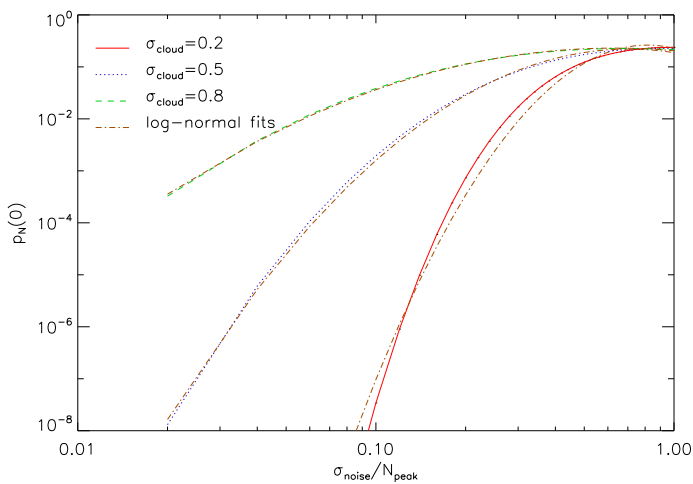}
      \caption{Demonstration of the log-normal behavior of the linear 
	zero column density PDF $p_N(0)$ as a function of the observational
	noise level. The three colored curves represent horizontal cuts through the
	surface shown in Fig.~\ref{fig_noise-zero-values} for different 
	values of the cloud PDF width. The brown dashed-dotted lines
	show the corresponding log-normal fits to the curves.}
         \label{fig_noise-zero-demo}
   \end{figure}

{\changed 
Closer inspection of the linear column density PDFs at $N=0$ for 
noise-affected cloud observations displayed in Fig.~\ref{fig_noise-zero-values}
shows that the noise dependence of $p_N(0)$ can also be approximated 
by some log-normal function. For every width of the original cloud PDF,
$\sigma_{\eta,{\rm cloud}}$, we can fit the zero column PDF as a function 
of the noise level by
\begin{equation}
p_N(0)= \frac{a_0}{N\sub{peak}}
\exp \left(-\frac{[\ln(\sigma\sub{noise}/N\sub{peak})-\ln(a_1)]^2}{2 a_2^2}
\right) 
\label{eq_eq_zero_pdf}
\end{equation}
in the range up to $\sigma\sub{noise}/N\sub{peak} = 1$. To avoid confusion
with the log-normal cloud column density PDFs, we use the less
intuitive parameter names, $a_0$, $a_1$, and $a_2$ here.\footnote{The log-normal
function for the noisy zero column PDF as a function of the noise 
level must not be confused with any of the log-normal functions discussed
in the main text of the paper. Those describe probabilities as a function 
of the (observed) column density.}.

The log-normal character is demonstrated in Fig.~\ref{fig_noise-zero-demo} showing 
three cuts through
Fig.~\ref{fig_noise-zero-values} in the $\sigma\sub{noise}/N\sub{peak}$
direction and the corresponding log-normal fits using Eq.~\ref{eq_eq_zero_pdf}. 
All curves in the plot show a clear parabola shape. Except for narrow
cloud widths, $\sigma_{\eta,{\rm cloud}} \la 0.2$, the representation of the 
curves by log-normals is very good. This allows us to condense the information
contained in the surface in Fig.~\ref{fig_noise-zero-values} into the
$\sigma_{\eta,{\rm cloud}}$ dependence of the three parameters $a_0$, $a_1$, and $a_2$
only.

% noise-zero-distribution
   \begin{figure}
   \includegraphics[angle=90,width=\columnwidth]{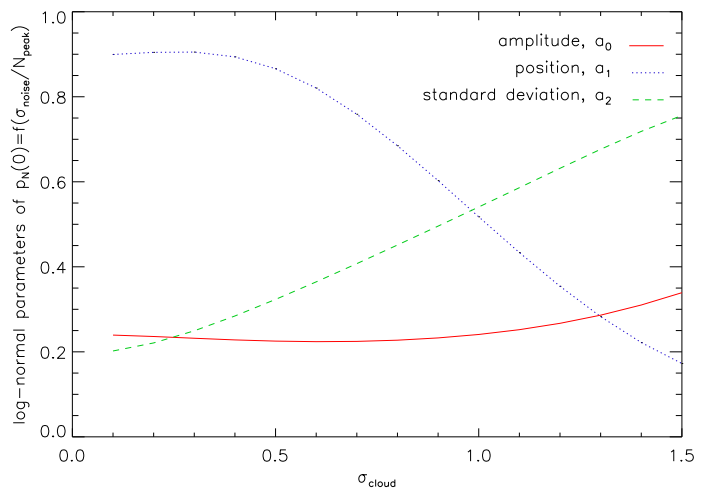}
      \caption{Parameters of the log-normal noise-amplitude dependence of 
	the zero-column density PDF ($p_N(0)$, Eq.~\ref{eq_eq_zero_pdf}) 
	shown as a function of the cloud PDF width. The values give the center, 
	$a_1$, amplitude, $a_0$, and width, $a_2$, of the 
	Gaussian on a logarithmic noise-amplitude scale in units of the 
	cloud PDF peak $N\sub{peak}$.}
         \label{fig_noise-zero-fit}
   \end{figure}

Figure~\ref{fig_noise-zero-fit} shows the parameters of the log-normals, 
$a_0, a_1, a_2$, for all horizontal cuts in Fig.~\ref{fig_noise-zero-values}
as functions of the cloud PDF width. 
As visible in Fig.~\ref{fig_noise-zero-demo},
the parabolas describing the zero-column density PDF become wider with
increasing cloud PDF width, the peak position moves to smaller 
columns, but the peak amplitude does not change significantly.
 
Using the three log-normal parameters, $a_0(\sigma_{\eta,{\rm cloud}}), 
a_1(\sigma_{\eta,{\rm cloud}}), a_2(\sigma_{\eta,{\rm cloud}})$,
plotted in Fig.~\ref{fig_noise-zero-fit}, we thus have a condensed
description of the zero-column PDF as a function of noise level 
and cloud PDF width from Fig.~\ref{fig_noise-zero-values}
that can be used to look up the noise level for any measured $p_N(0)$
and $\sigma_{\eta,{\rm cloud}}$.
}

\section{Recovering the PDF width of the contaminating cloud}
\label{appx_contamination}

The typical level of the contamination usually can be measured
by inspecting pixels at the boundaries of the cloud that are representative
of the contamination only. However, the number of those pixels
will be small compared to the whole map so that higher moments and
even the width of the contaminating PDF may not be known with sufficient 
accuracy. However, in principle one can infer the width by inspecting the
zero-column-density of the offset-corrected PDF, equivalent 
to our approach of measuring the noise in Sect.~\ref{sect_zero_pdf}. 
The colors at the lower edge of Fig.~\ref{fig_pdfscan_contam_subtracted}
demonstrate that for high contamination levels,
i.e. when the reconstructed PDF is too wide compared to the cloud PDF, 
it extends to low column densities including $p_N(0)$. A relation 
between contamination width, $\sigma\sub{contam}/\sigma_{\eta,{\rm cloud}}$ 
and the zero-column PDF $p_N(0)$ then could be used to deduce the
unknown contamination width. 

   \begin{figure}
   \includegraphics[angle=90,width=\columnwidth]{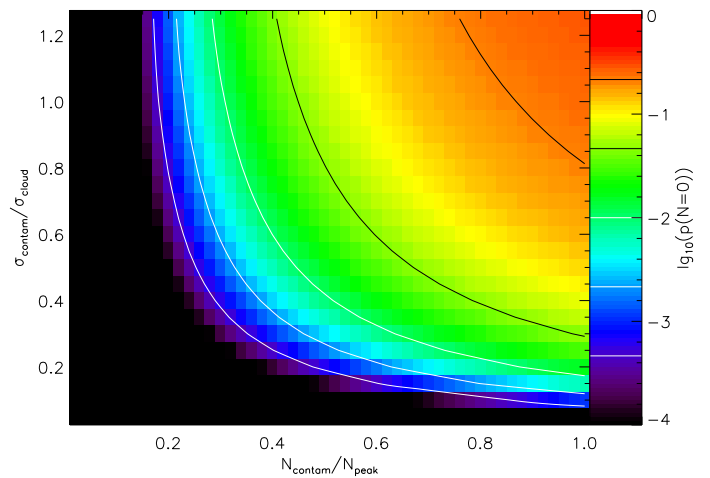}
      \caption{Linear column density PDF after offset correction at
          zero column, $p_N(0)$, as a function of the amplitude of the
	contamination and the width of the contaminating PDF. Colors
    	and contours are given on a logarithmic scale.} 
         \label{fig_contam_subtracted_zero}
   \end{figure}
   
In Fig.~\ref{fig_contam_subtracted_zero} we show the linear column
density $p_N(0)$ measured after the subtraction of the typical
contamination level as a function of the relative amplitude and 
width of the contaminating PDF. We can accurately measure the width 
of the contaminating PDF from the plot if it shows a steep gradient
of $p_N(0)$ in the direction of the contamination width, i.e.
in the vertical cut through the plot given by the known contamination
amplitude $N\sub{contam}$. Unfortunately, we find such a steep gradient
only in case of relatively narrow PDFs of the contaminating structure of
less than $\sigma\sub{contam} < 0.5 \sigma_{\eta,{\rm cloud}}$, i.e. in 
a regime where the recovery of the cloud PDF by the constant offset
correction anyway does not need any correction, but a 
relatively shallow behavior for broader contaminations.

   \begin{figure}
   \includegraphics[angle=90,width=\columnwidth]{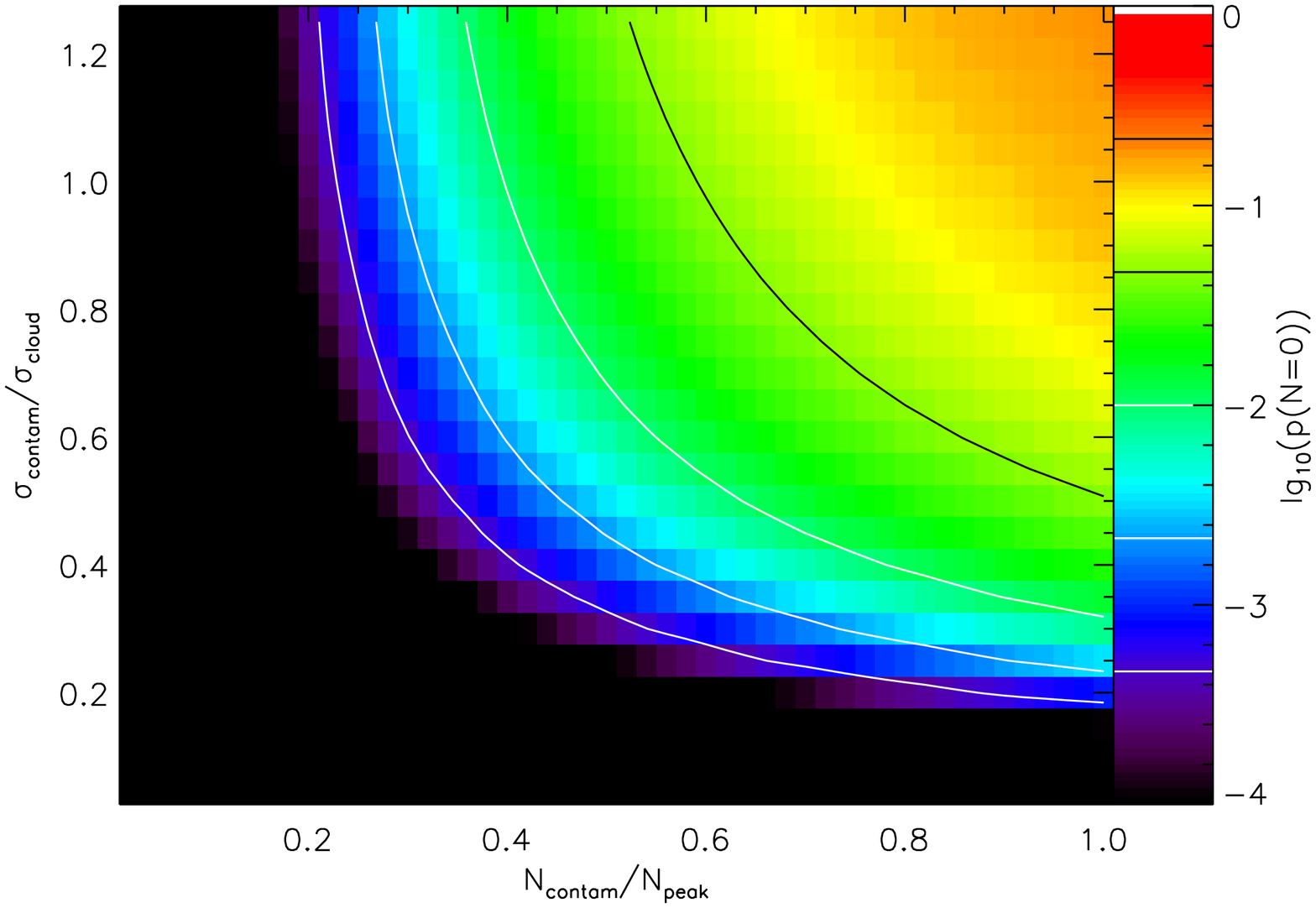}
	\vspace{0.3cm}\\
   \includegraphics[angle=90,width=\columnwidth]{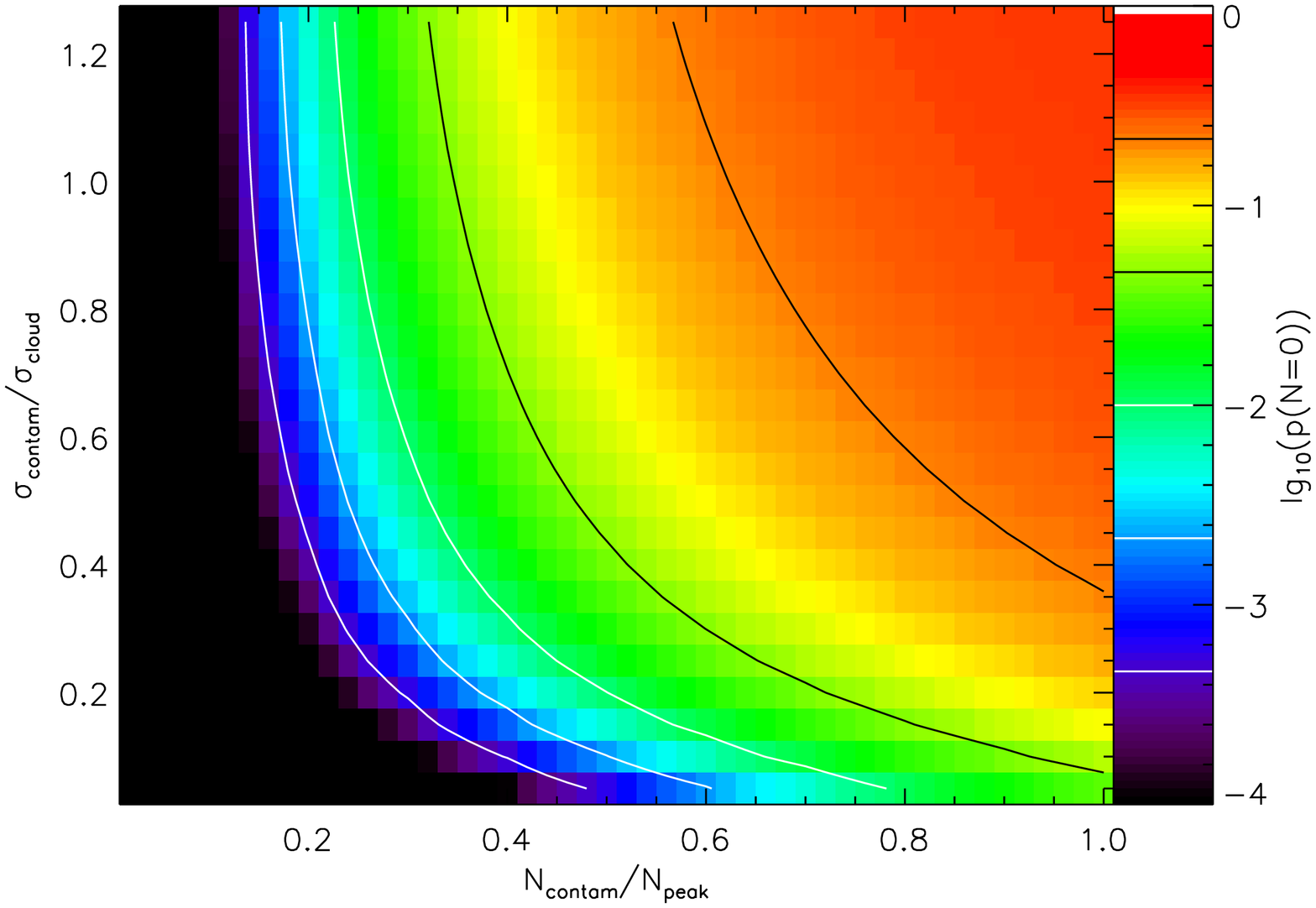}
   \caption{Same as Fig.~\ref{fig_contam_subtracted_zero} but
      using an imperfect subtraction of the typical contamination
      level. In the upper plot we assume that the contamination level is
      underestimated by 20\,\%, in the lower plot that it is 
       overestimated by 20\,\%.}
      \label{fig_contam_fuzzy_subtracted_zero}
   \end{figure}
   
Moreover, this approach depends on the accurate knowledge of the
contamination level, $N\sub{contam}$. Any ``overcorrection'' of the contamination
will shift the PDF towards lower densities, strongly increasing the
zero column density level. A small shift can lead to a significant 
change of the measured probability $p_N(0)$.
In Fig.~\ref{fig_contam_fuzzy_subtracted_zero} we show the
resulting zero-column PDF when assuming a 20\,\% error in the
amplitude of the contamination correction. 
When comparing the two plots in Fig.~\ref{fig_contam_fuzzy_subtracted_zero} 
we see that the zero-column PDF can change by
a factor of 100 in the sensitive range, while it still changes
by a factor of five in the range of the weak dependence from
the width of the contaminating PDF. 
Hence, it is practically very difficult to use the zero-column PDF 
after the line-of-sight correction to estimate the width of the 
PDF of the contaminating structure.
A direct measurement from a field tracing only the contamination is
always preferable.

\end{document}